\documentclass[useAMS, twocolumn, usenatbib]{mn2e}
\usepackage{graphicx}
\usepackage{amsmath, amssymb}
\usepackage{dcolumn}
\usepackage{bm}
\usepackage{soul}

\title{Consequences of Strong Compression in Tidal Disruption Events}
\author[N.~Stone, R.~Sari and A.~Loeb]{Nicholas Stone$^1$\thanks{E-mail:
nstone@cfa.harvard.edu}, Re'em Sari$^2$, and Abraham Loeb$^1$
\\$^1$Astronomy Department, Harvard University, 60 Garden St.,
Cambridge, MA 02138, USA
\\$^2$ Racah Institute for Physics, The Hebrew University, Jerusalem 91904, Israel}

\begin{document}

\date{\today}
\maketitle

\begin{abstract}

The tidal disruption of a star by a supermassive black hole (SMBH) is a highly energetic event with consequences dependent on the degree to which the star plunges inside the SMBH's tidal sphere.  We introduce a new analytic model for tidal disruption events (TDEs) to analyze the dependence of these events on $\beta$, the ratio of the tidal radius to the orbital pericenter.  We find, contrary to most previous work, that the spread in debris energy for a TDE is largely constant for all $\beta$.  This result has important consequences for optical transient searches targeting TDEs, which we discuss.  We quantify leading-order general relativistic corrections to this spread in energy and find that they are small.  We also examine the role of stellar spin, and find that a combination of spin-orbit misalignment, rapid rotation, and high $\beta$ may increase the spread in debris energy.  Finally, we quantify for the first time the gravitational wave emission due to the strong compression of a star in a high-$\beta$ TDE.  Although this signal is unlikely to be detectable for disruptions of main sequence stars, the tidal disruption of a white dwarf by an intermediate mass black hole can produce a strong signal visible to Advanced LIGO at tens of megaparsecs.

\end{abstract}

\section{Introduction}
Stars which pass too close to supermassive black holes (SMBHs) are disrupted by the enormous gravitational gradients acting on them.  The eventual fallback of $\sim 50\%$ of the star's mass onto the black hole can produce a highly luminous, multiwavelength flare - the primary observable signature of a tidal disruption event (TDE).  Over a dozen strong TDE candidates have been observed, with most detections made in X-ray \citep{Bade+96, KomossaGreiner99, Gezari+03} or UV \citep{Gezari+06, Gezari+08, Gezari+09}, but some in optical archival \citep{vanVelzen+11} and transient \citep{Cenko+12, Gezari+12} searches.  Recently, collimated jets from two relativistic TDE candidates have been detected by the Swift satellite \citep{Levan+11, Bloom+11, Zauderer+11, Cenko+11}.  

The rate of TDEs is highly uncertain, from both theoretical and observational perspectives.  On the observational side, uncertainties stem from both the low sample size and unclear sources of systematic error; nonetheless, observational estimates of the TDE rate per galaxy generally find $\dot{N}_{\rm TDE}\sim 10^{-4}-10^{-5}~{\rm yr}^{-1}$ \citep{Donley+02, Gezari+08}.  This is in rough agreement with the wide range of theoretical predictions for the TDE rate, which invoke different ways to scatter stars into the SMBH ``loss cone'' (the region of low angular momentum phase space containing orbits which pass inside the tidal sphere).  The most theoretically secure method of feeding stars into the loss cone is standard two-body relaxation, which sets a conservative lower limit on $\dot{N}_{\rm TDE}$ between $10^{-4}$ and $10^{-6}~{\rm yr}^{-1}$ \citep{FrankRees76, LightmanShapiro77, CohnKulsrud78, MagorrianTremaine99, WangMerritt04}.  Alternative mechanisms for enhancing the TDE rate include triaxiality in a galaxy's nuclear potential \citep{MerrittPoon04}, encounters with massive perturbers \citep{Perets+07}, the effect of an inspiraling secondary SMBH \citep{Ivanov+05, Chen+09, Chen+11, WeggBode11}, and gravitational wave recoil after the merger of a binary SMBH \citep{StoneLoeb11}.

Past theoretical work on TDEs has included analytic estimates of event energetics and timescales \citep{Rees88, Phinney89, Ulmer99, StrubbeQuataert09}, hydrodynamical simulations of the disruption process in both smoothed-particle-hydrodynamics \citep{NoltheniusKatz82, EvansKochanek89, Laguna+93, Lodato+09} and mesh \citep{Guillochon+09, GuillochonRamirez12} codes, and radiative transfer work to quantify emission and absorption processes in TDEs \citep{KasenRamirez10, StrubbeQuataert11}.  The large hierarchy of time and length scales involved in a TDE makes it difficult to self-consistently simulate one from disruption to the onset of accretion, so other work has focused on the formation \citep{Kochanek94} and evolution \citep{ArmijoPacheco11} of TDE accretion disks.  The properties of these flares depend crucially on the spread in orbital specific energy of the post-disruption debris streams, as that spread sets the mass fallback rate onto the SMBH.  All fallback rates are generally expected to produce multicolor blackbody emission from the accretion disks they feed \citep{LodatoRossi11}, while super-Eddington fallback rates may drive powerful outflows, increasing the optical luminosity by orders of magnitude \citep{LoebUlmer97, StrubbeQuataert09, StrubbeQuataert11}.

Much of the pioneering work on TDEs was done in the 1980s using the semi-analytic affine model \citep[hereafter CL83]{CarterLuminet83}, which treats the disrupting star as a set of concentric ellipsoidal shells evolving under the combined influences of self-gravity, pressure, and the SMBH tidal field \citep{CarterLuminet85, LuminetCarter86}.  This model has found a wide range of uses, and has been generalized to include both thermonuclear reaction networks \citep{LuminetPichon89} and general relativistic (GR) effects \citep{LuminetMarck85}, although its validity tends to break down at late times as the stellar debris exits the tidal sphere.  One key finding of the affine model is that during the early stages of disruption, prior to the star's arrival at pericenter, motion orthogonal to the orbital plane decouples from motion within, leading to a strong, one-dimensional compression (a vertical ``pancaking'') of the star.  This effect scales strongly with the penetration factor $\beta$, defined as the ratio of the tidal radius $R_{\rm t}$ to the pericenter radius $R_{\rm p}$.  The pancaking of the star is reversed by the buildup of internal pressure, which leads to a rebound in the vertical direction.  Shock formation accompanies this rebound (and occasionally the infall period prior to maximum compression), with X-ray shock breakout a potential though as yet undetected observational signature of TDEs \citep{Kobayashi+04, Guillochon+09}.

In this paper, after establishing basic dynamical features of TDEs (\S 2) we present a new analytic model to analyze the tidal free fall of the star prior to its maximum vertical compression (\S 3).  In many ways, this represents a simplification of the affine model, and its primary appeal is its greater analytic tractability.  Using our model we correct a longstanding error in the literature on the spread in debris energy $\Delta \epsilon$.  We verify the robustness of our estimates by considering redistribution of vertical collapse energy to in-plane motion (\S 4), the desynchronization of vertical collapse (\S 5), and leading-order GR corrections (\S 6), the latter of which are found to be small.  We examine the gravitational waves (GWs) generated by rapid changes in the star's quadrupole moment during maximum compression, and find them to be detectable by Advanced LIGO for disruptions of white dwarfs (\S 7).  We conclude with the observational implications of our work, which primarily involve the suppression of strongly super-Eddington TDEs (\S 8), and a general discussion (\S 9).

\section{Dynamical Energy Spread}
A star is tidally disrupted if the pericenter of its orbit, $R_{\rm p}$, lies inward of the tidal radius,
\begin{equation}
R_{\rm t}=R_*(M_{\rm BH}/M_*)^{1/3}.
\end{equation}
Here $M_*$ and $R_*$ are the mass and radius of the victim star, and $M_{\rm BH}$ is the black hole mass.  In reality, this expression for the tidal radius is not exact and contains weak, order unity dependences on stellar structure \citep{Diener+95}, stellar spin, and black hole spin \citep{Kesden11}.  We ignore these complications in this paper.  Very shortly after entry into the tidal sphere (and before pericenter passage for $\beta>1$), the SMBH's tidal forces do an amount of work exceeding the star's gravitational binding energy, and the star's fluid elements begin moving on roughly geodesic trajectories.  In the standard picture, hydrodynamic forces are subsequently neglected and the specific orbital energy $\epsilon$ of the debris streams is ``frozen in,'' with a spread given by
\begin{equation}
\Delta \epsilon = k\frac{GM_{\rm BH} R_*}{R_{\rm p}^2}, \label{UncorrectedEnergy}
\end{equation}
where $G$ is the gravitational constant and $k$ a constant of order unity related to stellar structure and rotation prior to disruption.  This approximate estimate can be obtained by Taylor-expanding the SMBH potential around the star at pericenter, or alternatively by multiplying the equivalent tidal acceleration at pericenter $A_{\rm p}\sim (GM_{\rm BH}/R_{\rm p}^2)(R_*/R_{\rm p})$ by the dynamical time $T_{\rm p}\sim (GM_{\rm BH}/R_{\rm p}^3)^{-1/2}$ to get $\Delta V_{\rm p}=A_{\rm p}T_{\rm p}$.  Using $V_{\rm p}=(2GM_{\rm BH}/R_{\rm p})^{1/2}$, one can then find $\Delta \epsilon=V_{\rm p}\Delta V_{\rm p} \sim GM_{\rm BH}R_*/R_{\rm p}^2$.  We note that Eq. \ref{UncorrectedEnergy} is widely used in the literature \citep{EvansKochanek89, Kochanek94, Ulmer99, KasenRamirez10, StrubbeQuataert09, LodatoRossi11}.  

 However, this reasoning is incorrect; by the time the star reaches pericenter its fluid elements are moving on almost ballistic trajectories.  As the star plunges into the tidal sphere, internal forces become subdominant to the SMBH tidal field, with the ratio of the tidal to the self-gravitational acceleration given by $a_{\rm t}/a_{\rm g}\approx (R_{\rm t}/R)^3$.  Here $R$ is the orbital separation.  The work done by internal forces decreases more slowly, $\sim GM_*R/R_*^2$, although this simple expression overestimates the amount of work done by internal forces, which at $R\approx R_{\rm t}$ will self-cancel each other to first order (given that the star is initially in hydrostatic equilibrium).  

To accurately evaluate $\Delta\epsilon$ at pericenter passage, one would need to account for distortions in the free-falling star's physical shape, as well as internal velocities.  At any point along the star's orbit, a Cartesian coordinate system \citep{BrassartLuminet08} will define the principal axes (eigenvectors) of the tidal tensor.  If we define $\hat{X}$ parallel to the vector connecting the star and SMBH, $\hat{Y}$ in the orbital plane but perpendicular to $\hat{X}$, and $\hat{Z}$ perpendicular to the orbital plane, the star will be stretched in the $\hat{X}$ direction but compressed in the $\hat{Y}$ and $\hat{Z}$ directions.  The compression along the $\hat{Y}$ axis reduces the potential gradient across the star, invalidating the above formula; further inaccuracy is introduced by the internal motions (i.e. velocity shear among ballistic debris trajectories) of the star within the SMBH's tidal sphere.  A more accurate estimate of the spread in specific energy can be found by taking the potential gradient at the moment of tidal disruption, i.e. when the star crosses the tidal sphere and becomes unbound, as after this point the motion of the debris becomes roughly geodesic.  This revision to the approximation of energy freeze-in yields
\begin{equation}
\Delta \epsilon = k\frac{GM_{\rm BH} R_*}{R_{\rm t}^2}. \label{CorrectedEnergy}
\end{equation}
We note that an analogous conclusion (on the $\beta$-independence of the energetics of tidal disruption) can be seen in tidal separations of binary stars by SMBHs \citep[hereafter SKR10]{Sari+10}.  We can alternatively use $\beta=R_{\rm t}/R_{\rm p}$ to rewrite Eqs. \ref{UncorrectedEnergy}, \ref{CorrectedEnergy} as
\begin{equation}
\Delta \epsilon = k\beta^n \frac{GM_{\rm BH}R_*}{R_{\rm t}^2}, \label{ParamEnergy}
\end{equation}
with $n=2$ for the standard, Eq. \ref{UncorrectedEnergy} picture and $n=0$ for our revised, Eq. \ref{CorrectedEnergy} analysis.  In the sections below, our more detailed analysis of the tidal compression experienced by the star will examine if intermediate or piecewise values of $n$ are more appropriate.  The observational implications of changes to $\Delta \epsilon$ are discussed in \S 8.

\section{Free Solutions, and Free Collapse}
Other factors could influence or invalidate the simple analytic argument presented in \S 2, such as redistribution of energy during the moments of maximum compression, GR corrections, stellar spin, or simply work done on the star's fluid elements by subdominant internal forces inside the tidal sphere.  In this section, we introduce a new analytic model for the tidal free fall of a disrupted star that will help us approach these issues.

Because the dominant source of TDEs is expected to be stars scattered onto radial orbits from $\sim$ pc scales \citep{MagorrianTremaine99, WangMerritt04}, we assume a parabolic orbit for the center of mass of the star, with distance from the SMBH given by:
\begin{equation}
R=\frac{2R_{\rm p}}{1+\cos f}.
\end{equation}
For such an orbit time $t$ is related to true anomaly $f$ via
\begin{equation}
t=\frac{1}{3}\left(\frac{2R_{\rm p}^3}{GM_{\rm BH}}\right)^{1/2}\tan(f/2)\left(3+\tan ^2(f/2)\right),
\end{equation}
although the differential form
\begin{equation}
\frac{df}{dt}=\frac{1}{8^{1/2}}(1+\cos f)^2\sqrt{\frac{GM_{\rm BH}}{R_{\rm p}^3}}
\end{equation}
is more generally useful.  We set $t=f=0$ at $R=R_{\rm p}$, and use $\dot{f}>0$ throughout this paper.

The sequence of events in a TDE, first noted by CL83, will be useful shorthand for us, so we introduce it here.  Phase I  (near-equilibrium) of a TDE lasts while $R>R_{\rm t}$, and the star remains in approximate if slightly perturbed equilibrium.  Phase II (free fall) begins when the star crosses the tidal sphere and becomes gravitationally unbound; in this paper we will treat the transition between Phases I and II as instantaneous, an assumption we justify below in \S 5.  The assumption of tidal free fall is very useful because of the existence of analytic, ``free'' solutions to the Hill equations in the parabolic restricted 3-body problem, but it is not immediately obvious that internal forces in the star can be neglected for $R<R_{\rm t}$.  To first order the approximation seems reasonable because the ratio of tidal acceleration to self-gravitational acceleration grows quickly, as $a_{\rm t}/a_{\rm g}\approx (R_{\rm t}/R)^3$ for the bulk of the star.  Furthermore, the star's internal pressure and self-gravity partially cancel each other, further reducing their combined contribution.  For now, we assume the validity of the free fall assumption, but after developing further machinery we will justify it further in \S 5.

During this free fall, the star is compressed perpendicular to the orbital plane (along $\hat{z}$) and in one direction within the orbital plane, while being stretched along the other in-plane direction.  Although for the limiting case of radial infall the problem is self-similar in all three dimensions (SKR10), the rotation of the line connecting the star's center of mass to the SMBH breaks the in-plane similarity.   By the time the star has reached pericenter, the $\hat{x}$ direction (which is parallel to the line between the SMBH and the orbital pericenter), is compressed, and $\hat{y}$ is stretched, but the distortions are both much smaller than the compression orthogonal to the orbital plane.  Shortly after passing pericenter, synchronous tidal free fall in the $\hat{z}$ direction leads to very strong compression of the star, which is eventually reversed by hydrodynamic forces.  Phase III (bounce) begins when hydrodynamical forces become strong enough to begin slowing the collapse of the star along its vertical axis.  Once the star's collapse has reversed, hydrodynamical forces quickly become negligible again, and Phase IV (the rebound) begins, with stellar gas once again moving on ballistic trajectories.

We take as initial conditions for Phase II a spherically symmetric star at the tidal sphere, with fluid elements possessing initial positions $\vec{r}$ (the coordinate origin tracks the star's center of mass) and initial velocities in the center of mass frame $\vec{u}(\vec{r})$.  Making the approximation that upon entering the tidal sphere, internal forces become negligible unless and until compression triggers shock formation or isentropic pressure buildup, we take the pre-shock trajectories of these fluid elements to be completely ballistic.  This means that their trajectories are given by the parabolic Hill equations, $\vec{r}_{\rm H}=\{x_{\rm H}, y_{\rm H}, z_{\rm H}\}$.  The free solutions to these equations, neglecting self-gravity, can be written in closed form (SKR10) using coordinates where distance has been normalized by $R_{*}$ and time by $\sqrt{R_{*}^3/(GM_{*})}$; we denote such coordinates in this paper by writing tildes over them.  All other coordinates are in physical units, unless otherwise noted.  Although there are 6 independent solutions to these equations, motion out of the orbital plane is decoupled from motion within it, so only two are relevant for perturbed motion in the $\hat{z}$ direction:
\begin{align}
\tilde{z}_{\rm H}=E \tilde{z}_{\rm E} + F \tilde{z}_{\rm F} \notag \label{ansatz} \\
\tilde{z}_{\rm E}=\frac{1}{\beta}\frac{2\sin f}{1+\cos f}  \\
\tilde{z}_{\rm F}=\frac{1}{\beta}\frac{2\cos f}{1+\cos f}. \notag
\end{align}
Here $E$ and $F$ are undetermined coefficients that are set by the initial conditions described above.  In particular, if we require that a fluid element of initial position $z=z_0$ has initial velocity $w=\dot{z}=0$ at $f=f_{\rm t}$, where the true anomaly upon entry into the tidal sphere is given by 
\begin{equation}
f_{\rm t}=-\arccos(2/\beta-1), \label{ft}
\end{equation}
then
\begin{align}
&E=-\tilde{z}_0 \sqrt{\beta-1} \\
&F=\tilde{z}_0. 
\end{align}
If we introduce a tidal potential $\Psi$ felt in the rest frame of the star, the tidal acceleration is given by SKR10 as
\begin{equation}
\ddot{\tilde{z}}=-\frac{\partial \Psi}{\partial \tilde{z}}=-\beta^3 \frac{(1+\cos f)^3}{8}\tilde{z}. \label{selfSimilar}
\end{equation}
We note that the self-similarity of Eq. \ref{selfSimilar} implies that the free solutions all collapse to $z=0$ simultaneously at a true anomaly $f_{\rm c}$, although physically this collapse will be reversed shortly before by the buildup of pressure gradients strong enough to counteract the tidal forces compressing the star.  However, it is useful to solve for $f_{\rm c}$ using Eq. \ref{ansatz}:
\begin{equation}
\tan f_{\rm c} = \frac{1}{(\beta-1)^{1/2}}. \label{fc}
\end{equation}
From this formula we see that in the limit of $\beta \to \infty$, collapse along the z-axis occurs at $f_{\rm c}=0$, i.e. at pericenter, while in the marginal disruption limit of $\beta \to 1$, collapse occurs at $f_{\rm c}=\pi /2$, i.e. at a fixed point past pericenter.  We have already mentioned that the free solutions become less valid for small $\beta$ due to the increasing importance of internal forces, but we can see from Eq. \ref{fc} a second, stronger, inconsistency at low $\beta$, which  is that the free solutions dictate vertical collapse after the disrupted star leaves the tidal sphere, i.e. $f_{\rm c}>|f_{\rm t}|$.  This occurs for $\beta \lesssim 1.3$.

Although the onset of Phase III is dictated by compression in the $\hat{z}$ direction, the outcome of the bounce will be affected by motion within the orbital plane during Phase II, when $f<f_{\rm c}$.  We therefore describe here the free solutions within the orbital plane (SKR10):

\begin{align}
\tilde{x}_{\rm H}=A\tilde{x}_{\rm A}+B\tilde{x}_{\rm B}+C\tilde{x}_{\rm C}+D\tilde{x}_{\rm D} \notag \\
\tilde{y}_{\rm H}=A\tilde{y}_{\rm A}+B\tilde{y}_{\rm B}+C\tilde{y}_{\rm C}+D\tilde{y}_{\rm D} \notag \\
\tilde{x}_{\rm A}=-\frac{1}{\beta}\frac{\sin{f}}{1+\cos{f}} \notag \\
\tilde{y}_{\rm A}=\frac{1}{\beta}\frac{\cos{f}}{1+\cos{f}} \notag \\
\tilde{x}_{\rm B}=-\frac{1}{\beta}\sin{f} \notag \\
\tilde{y}_{\rm B}=\frac{1}{\beta}(1+\cos{f})\label{xyFreeSols}  \\
\tilde{x}_{\rm C}=\frac{1}{\beta}(2-\cos{f}) \notag \\
\tilde{y}_{\rm C}=-\frac{1}{\beta}\cos{f}\tan(f/2) \notag \\
\tilde{x}_{\rm D}=\frac{1}{\beta}(8+12\cos{f})\tan^4(f/2) \notag \\ 
\tilde{y}_{\rm D}=\frac{1}{\beta}\frac{35\sin{f}-2\sin(2f)+3\sin(3f)}{(1+\cos{f})^2} \notag
\end{align}

If we consider a point on the star with an initial position, relative to the star's center of mass, of $(x_0, y_0, z_0)$ and zero initial velocity (here, as before, ``initial'' refers to $f=f_{\rm t}$, i.e. crossing into the tidal sphere), then we have 4 initial conditions for 4 unknowns: $\{A, B, C, D\}$.  Using Eq. \ref{ft}, we find

\begin{align}\label{xyICs}
A=&\frac{1}{\beta^2} \left(-8 \tilde{x}_0 \sqrt{\beta -1} + 2\tilde{y}_0 (\beta^2+2\beta -4)\right) \\ 
B=&\frac{1}{5\beta^2} \Big(2 \tilde{x}_0 \sqrt{\beta -1} (\beta^3-4\beta^2+8) \\
&+ \tilde{y}_0 (9\beta^3-12\beta^2-8\beta+16)\Big) \notag \\
C=&\frac{1}{\beta^2} \left(\tilde{x}_0 (2\beta^2+\beta-2) - 2 \tilde{y}_0 \sqrt{\beta-1} (\beta^2 -1)\right) \\
D=&-\frac{1}{20\beta^2} \left(\tilde{x}_0 (\beta - 2) + 2 \tilde{y}_0 \sqrt{\beta -1}\right)
\end{align}
All six of the free solutions we have listed can be thought of as slight perturbations to different orbital elements of the parabolic center of mass trajectory, boosted into the center of mass frame.

\begin{figure}
\includegraphics[width=85mm]{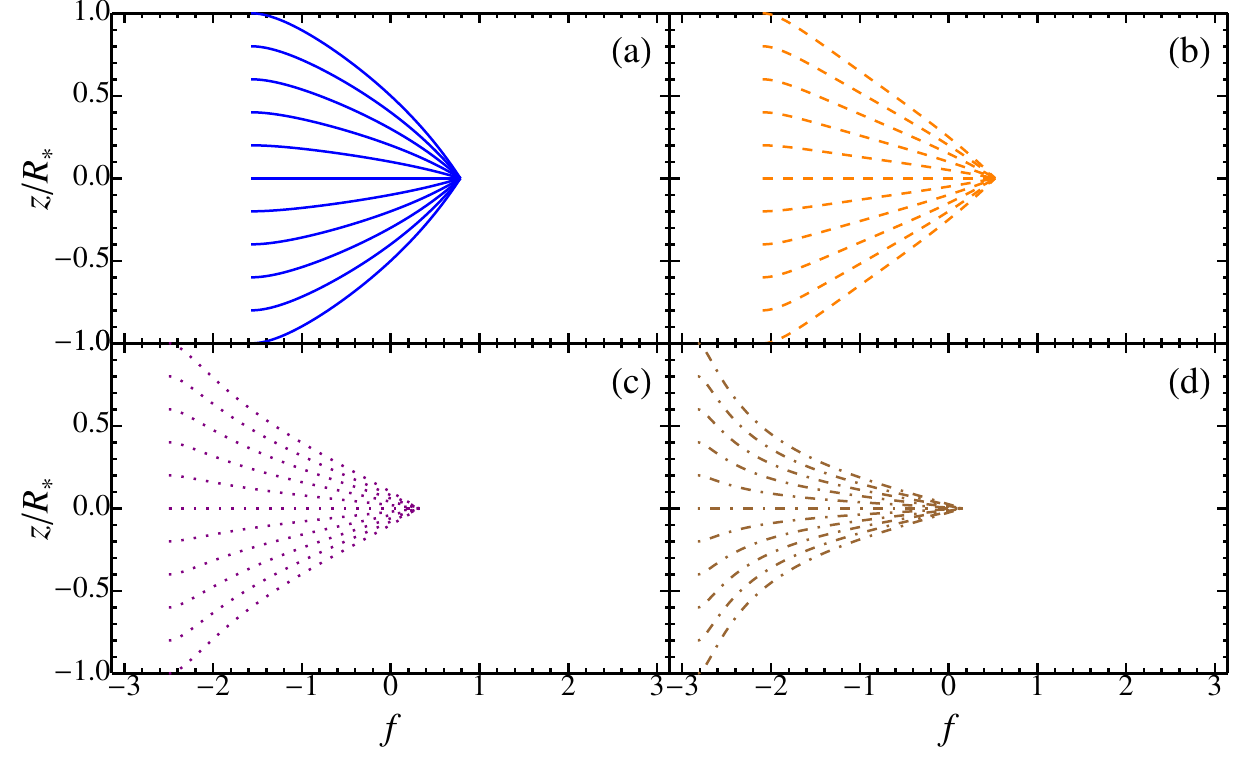}
\caption{Normalized height $\tilde{z}=z/R_*$ versus true anomaly $f$ for the vertical collapse of one-dimensional stars at varying $\beta$.  The solid blue curves in panel (a) are $\beta=2$; the dashed orange curves in panel (b) are $\beta=4$; the dotted purple curves in panel (c) are $\beta=10$; the dot-dashed brown curves in panel (d) are $\beta=40$.  Each scenario is initialized at $f=f_{\rm t}(\beta)$.  Note that $f=0$ corresponds to pericenter.}
\label{zFreeFig}
\end{figure}

\begin{figure*}
\includegraphics[width=180mm]{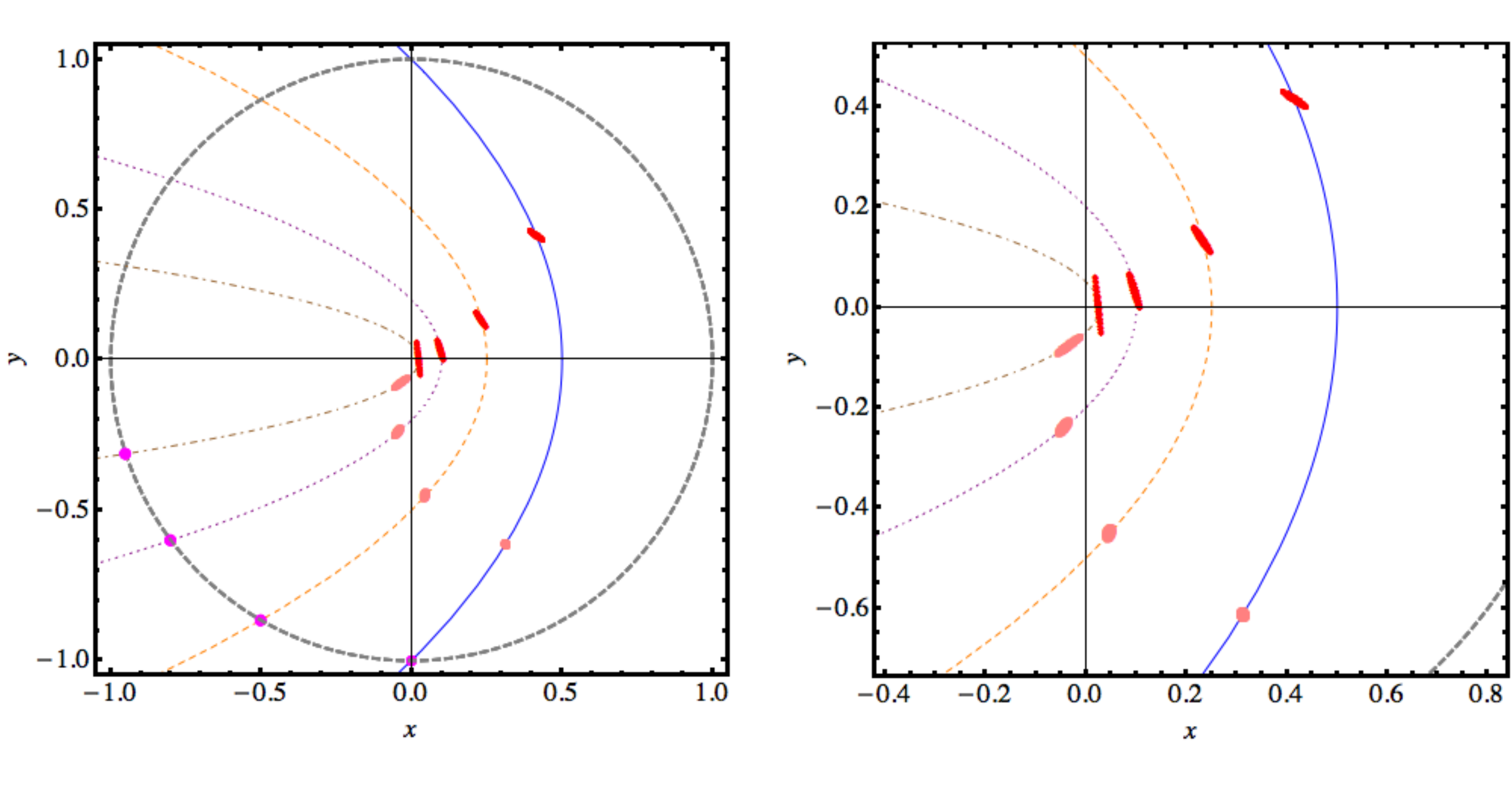}
\caption{The $x$ and $y$ coordinates (in units of $R_{\rm t}$, which for the $10^6M_{\odot}$ SMBH in this example is 100 $R_{\odot}$) of the free solutions for varying $\beta$, translated so that the origin lies on the SMBH.  As before, we mark the $\beta=2$ trajectory as solid blue, $\beta=4$ as dashed orange, $\beta=10$ as dotted purple, and $\beta=40$ as dot-dashed brown.  The free solutions for an initially circular midplane slice of a star are magenta at $f=f_{\rm t}$, pink at $f=0.7f_{\rm t}$, and red at $f=f_{\rm c}$.  The tidal radius is marked as a gray dashed circle.  The right plot is a zoomed-in version of the left.  The free solutions are breaking down for the $\beta=40$ curve near pericenter, as the long axis of the star exceeds the orbital radius in size.}
\label{xyFreeFig}
\end{figure*}

We have now exactly specified the motion of the idealized star's fluid elements in the orbital plane during Phase II.  We plot the vertical free solutions for a variety of $\beta$ in Fig. \ref{zFreeFig}, and snapshots from motion within the orbital plane in Fig. \ref{xyFreeFig}.  Here we list several important features of the free solutions, when they are initialized with static spheres of matter at $f=f_{\rm t}$:
\begin{itemize}
\item For $f>f_{\rm t}$, an initially spherical shell of matter will deform into a sequence of roughly ellipsoidal shapes.  It is simple to demonstrate that they do not generally take the form of true ellipsoids, however.
\item Initially concentric spherical shells of matter remain concentric, with in-plane principal axes that remain aligned with those of other concentric shells.
\item Slices of the star through the orbital plane (z=0) maintain reflection symmetry across their rotating in-plane principal axes.
\item The derivation of the free solutions assumes that $R_*/R \ll 1$ (SKR10).  If we neglect stretching of the star, this is equivalent to requiring $\beta \ll (M_*/M_{\rm BH})^{1/3}$, a condition that is in general easily satisfied: a $10^6 M_{\odot}$ SMBH, if non-spinning, cannot disrupt solar-type stars with $\beta \gtrsim 11$ (they will plunge directly into the horizon).  Even a maximally spinning SMBH of this mass cannot disrupt solar-type stars with $\beta \gtrsim 47$.  The effects of tidal stretching will make it somewhat harder to satisfy this assumption, but only for the minority of the star's mass that is strongly stretched.
\end{itemize}

\begin{figure}
\includegraphics[width=85mm]{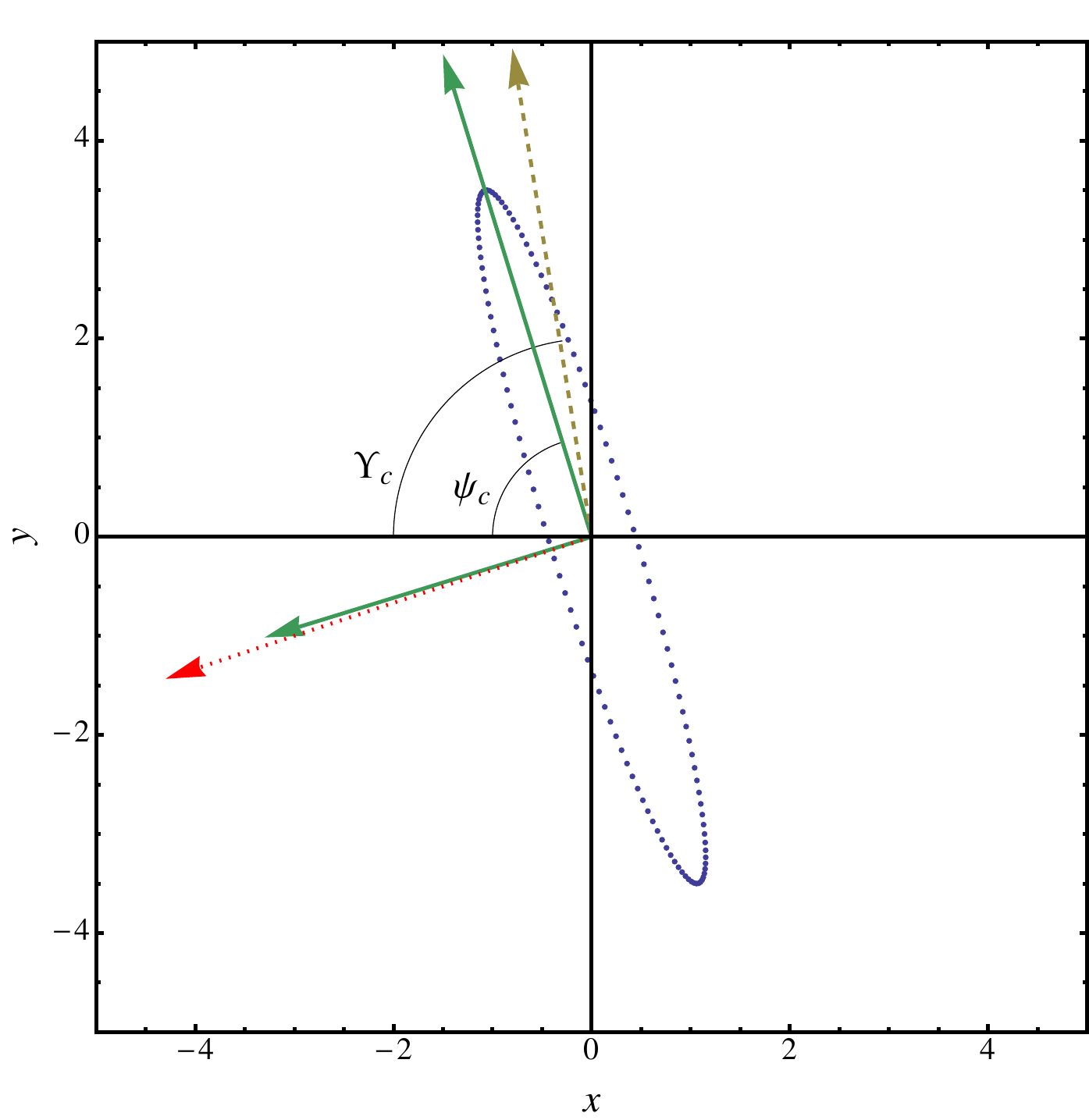}
\caption{An initially ($f=f_{\rm t}$) circular ring of stellar fluid elements has been tidally distorted by the time it reaches $f=f_{\rm c}$.  The principal axes of the distorted, free-falling body are the solid green vectors, the center of mass velocity is the dashed yellow vector, and the direction to the SMBH is the dotted red vector.  The angle $\psi_{\rm c}$ ($\Upsilon_{\rm c}$) is measured between the negative $\hat{x}$ direction and the long principal axis (stellar velocity vector).  We define the misalignment angle $\nu_{\rm c}=\Upsilon_{\rm c}-\psi_{\rm c}$.}
\label{angles}
\end{figure}

The free solutions allow us to directly solve for the stellar axis ratio as a function of $f$ or $t$, and it is trivial to do so numerically, but there is an exact analytic solution as well.  If we denote the lengths of the long and short principal axes of our tidally distorted star (within the orbital plane) as $r_{\rm long}$ and $r_{\rm short}$, respectively, we can solve for them by rewriting $x_0=\cos \theta _0$, $y_0=\sin \theta _0$, and finding the appropriate $\theta _0$.  More specifically, we set $\frac{\rm d}{{\rm d}\theta}R_{\rm H}^2(f)=0$ (with $R_{\rm H}^2=x_{\rm H}^2+y_{\rm H}^2$), and solve for $\theta_{\rm ex}$, the values of $\theta_0$ which extremize $R_{\rm H}$.  More physically, we are searching for the initial angles $\theta_{\rm ex}$ around the star which at a later orbital phase $f$ will correspond to its principal axes in the orbital plane.  Once we have the initial angular positions of the principal axes, $\theta_{\rm ex}$, we can plug in to Eqs. (\ref{xyFreeSols}) and solve for the size of the principal axes at a later true anomaly $f>f_{\rm t}$.  We also find the misalignment angle $\nu$ between the long in-plane principal axis and the orbital velocity vector.  The in-plane stellar geometry is presented in Fig. \ref{angles}.

The algebra involved in this solution is unenlightening, so we leave the general solution $\theta_{\rm ex}(f)$ for numerical work and only derive analytic expressions for $\theta_{\rm ex}(f_{\rm c})$, which is the situation of greatest interest.  The details are contained in Appendix A, but we plot the results below in Fig. \ref{rShortLong}.  Specifically, these are the sizes of the principal axes at $f=f_{\rm c}$.  For comparison we plot curves of the high $\beta$ limiting behavior, for which $\tilde{r}_{\rm long} \approx \frac{4}{5}\beta^{1/2}+\frac{22}{5}\beta^{-1/2}$ and $\tilde{r}_{\rm short} \approx 2\beta^{-1/2}-\frac{23}{2}\beta^{-3/2}$.  
\begin{figure}
\includegraphics[width=85mm]{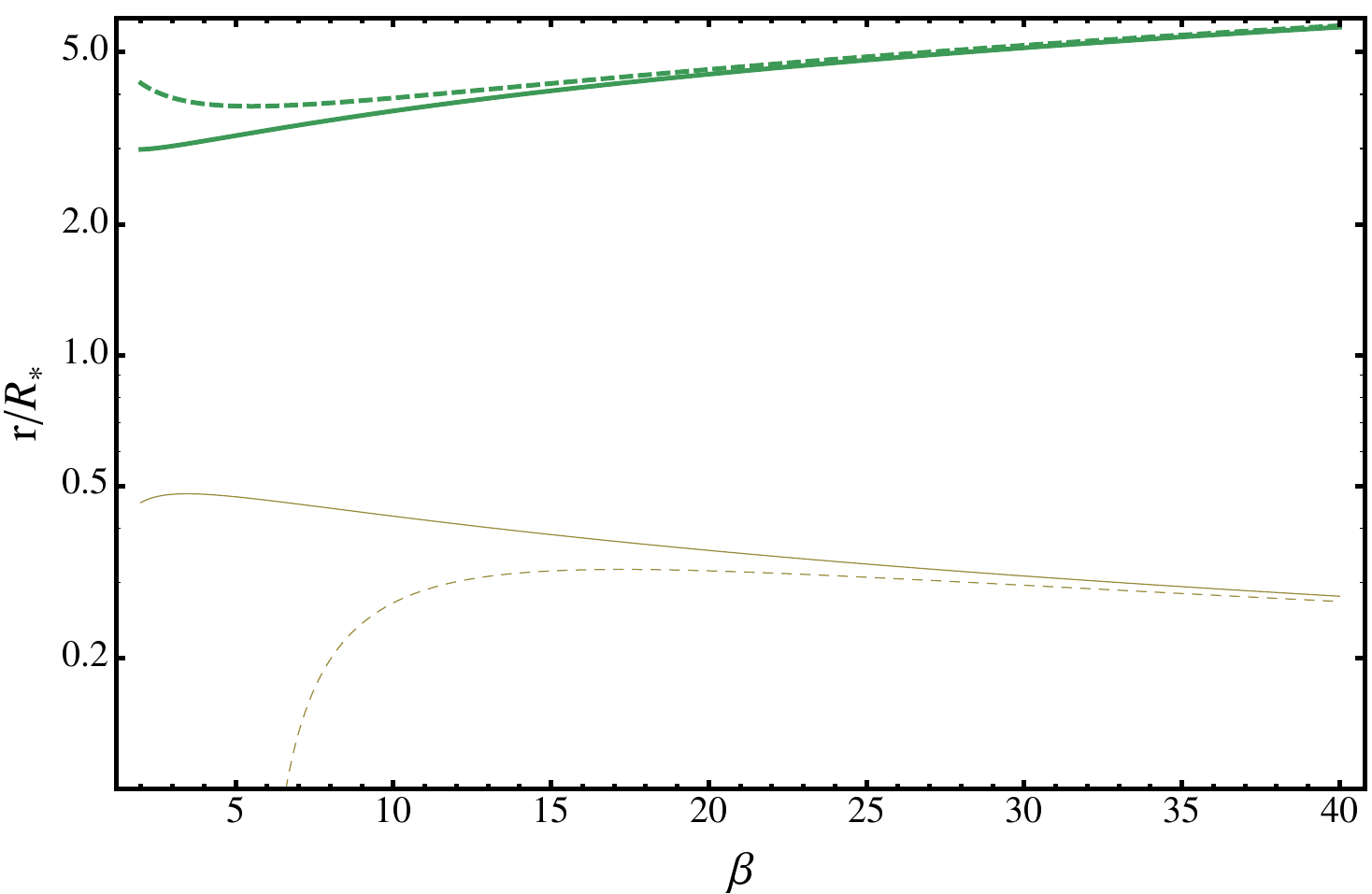}
\caption{The principal axis lengths, $r_{\rm long}$ and $r_{\rm short}$, of the distorted star (at $f=f_{\rm c}$) vs $\beta$.  Here $r_{\rm long}$ is plotted as thick green curves; $r_{\rm short}$ as thin yellow curves.  The exact solutions are solid lines, while the dashed curves are the high-$\beta$ Taylor expansions given by Eqs. (\ref{rLongLim}), (\ref{rShortLim}).}
\label{rShortLong}
\end{figure}

The primary interesting feature of the axis ratio calculations is that for disruptions of stars by supermassive black holes, the physically relevant range of $r_{\rm long}$ and $r_{\rm short}$ is quite narrow, being confined between 3 to 5 for the former, and 0.3 to 0.5 for the latter.  For the tidal disruption of a star by an intermediate mass black hole (IMBH), or perhaps more exotic disruption scenarios, a larger range of $\beta$ (and therefore $r_{\rm long}$, $r_{\rm short}$) can be attained, but for star-SMBH TDEs only a surprisingly narrow range of principal axis lengths are accessible.  This implies that the naive Taylor expansion of the SMBH potential as the star passes through $R=R_{\rm p}$, i.e. Eq. \ref{UncorrectedEnergy}, will fail primarily because of internal velocities within the free-falling stellar debris, and only secondarily because of distortions in the star's shape.

\begin{figure}
\includegraphics[width=85mm]{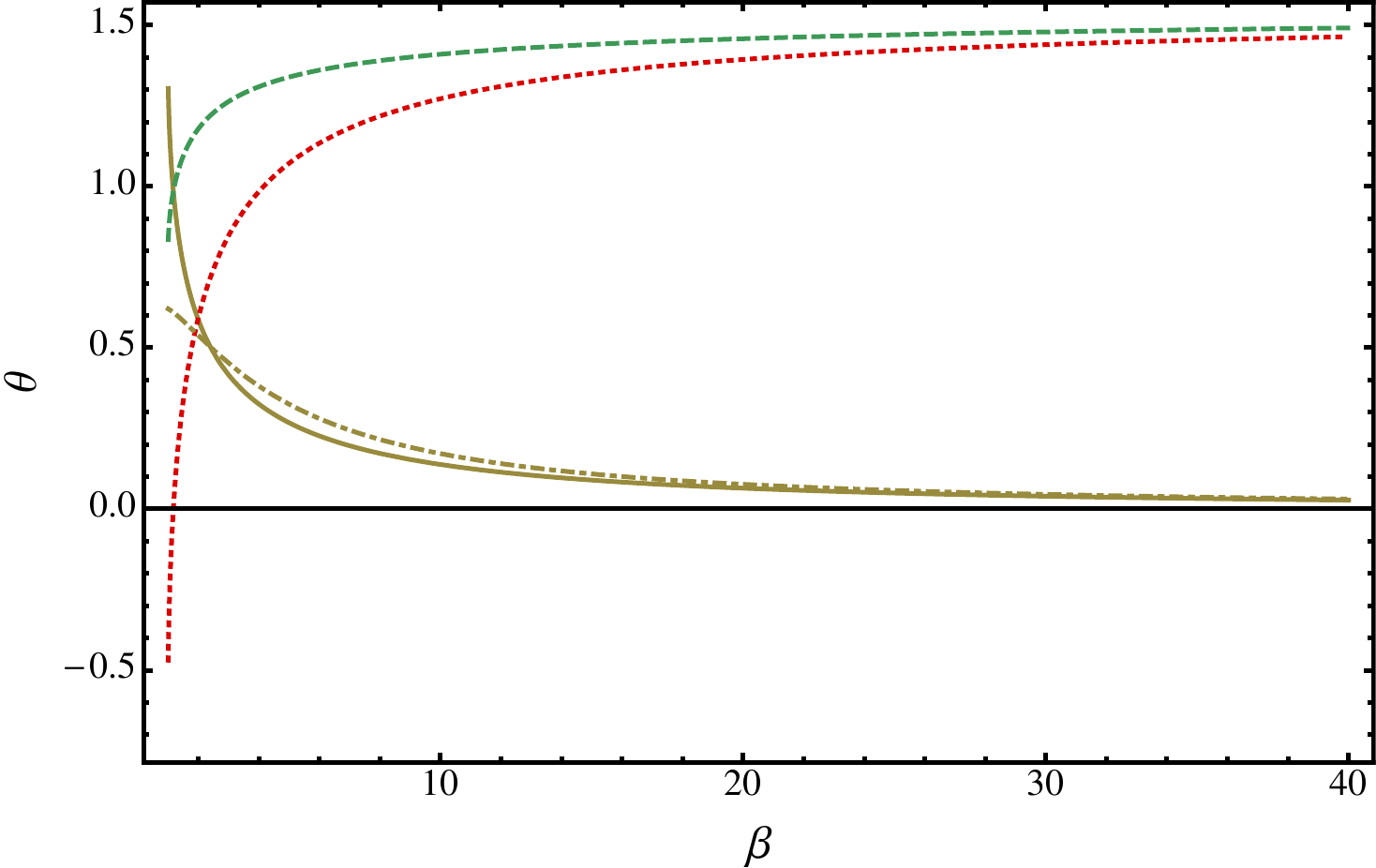}
\caption{Curves illustrating misalignment of the tidally raised bulge and the orbital velocity vector at the point of z-collapse ($f=f_{\rm c}$).  The red dotted curve is the angle between the negative x-axis and the tidal bulge ($\psi_{\rm c}$), the green dashed curve is the angle between the negative x-axis and the orbital velocity vector, and the solid yellow curve is the difference between them, i.e. the misalignment angle ($\nu_{\rm c}$).  These angles are plotted against the penetration factor $\beta$.  The high $\beta$ limit for $\nu_{\rm c}$ is the dot-dashed yellow curve.}
\label{misalignment}
\end{figure}

As we shall see in \S 4, when estimating energy redistribution during maximum vertical compression, the misalignment angle $\nu$ plays a larger role than the slowly-varying axis ratio.  This will prove relevant when calculating corrections to $\Delta \epsilon$, and is shown in Fig. \ref{misalignment}.  The angle $\nu_{\rm c}$ (as elsewhere, the subscript ${\rm c}$ denotes evaluation at $f=f_{\rm c}$) is found to be a rapidly decreasing function of $\beta$; to leading order, $\tan\nu_{\rm c} \propto \beta^{-3/2}$.

We can now use the free solutions $\{A, B, C, D, E, F\}$ to quantify precisely the spread in debris energy at the tidal radius.  Because these solutions can be thought of physically as perturbations to the orbital elements of a parabolic trajectory, all possess exactly zero energy except for the in-plane ``D'' solution, which has specific energy given by 
\begin{equation}
\epsilon=-\frac{20GM_*D}{R_*}\left( \frac{M_{\rm BH}}{M_*} \right)^{1/3} \beta,
\end{equation} 
where $D$ is the coefficient of the fourth in-plane free solution, corresponding to slight variations in the eccentricity of a near-parabolic orbit (SKR10).  If we initialize our free solutions with an unperturbed sphere, i.e. Eqs. \ref{xyICs}, then we find a specific energy for each fluid element of
\begin{equation}
\epsilon_u=\frac{GM_{\rm BH}R_*}{R_{\rm t}^2} (\tilde{x}_0(1-2/\beta)+2\tilde{y}_0\sqrt{\beta^{-1}-\beta^{-2}}) \label{epsilonU},
\end{equation}
where $\tilde{x}_0$ and $\tilde{y}_0$ are the initial positions of a debris stream relative to the star's center of mass at $R=R_{\rm t}$ normalized by the stellar radius.  Notably, the specific energy is to leading order independent of $\beta$, with the weak $\beta$-dependence becoming negligible at high penetration factors.    Defining $\tilde{x}_0=\tilde{r}_0 \cos \phi_0$ and $\tilde{y}_0=\tilde{r}_0 \sin \phi_0$, we can analytically extremize Eq. \ref{epsilonU} with respect to $\phi_0$, to find that the spread in energy of these unperturbed free solutions is actually fully independent of $\beta$: 
\begin{equation}
\Delta\epsilon_{\rm u}=\frac{2GM_{\rm BH}R_*}{R_{\rm t}^2} \label{DeltaEpsilonU}.
\end{equation}
In our idealized model of a spherical, stationary star undergoing tidal free fall, the assumption of energy freeze-in at the moment of disruption implies $n=0$, and, surprisingly, not even the weak $\beta$ dependence one might expect from Eq. \ref{epsilonU}.

In the following three sections, we examine the robustness of this model, and consider possible corrections to our expressions for $\Delta \epsilon$.  With limited exceptions, we find that the arguments made in this section remain generally valid.

\section{Total Vertical Collapse and Bounce}
For $f\approx f_{\rm c}$, motion in the vertical direction has decoupled from in-plane motion and the star undergoes a homologous vertical collapse.  In this regime, the vertical velocity of the free solutions near the point of maximum collapse is very close to a constant value, with $w_{\rm c} \propto \beta$, a result known since CL83, although the exact value, for arbitrary $f_{\rm c}$, is
\begin{equation}
w_{\rm c}=\beta \tilde{z}_0 \left( \frac{GM_*}{2R_*} \right)^{1/2}\left( (1-\beta^{-1})^{1/2} + 1  \right) \label{uc}.
\end{equation}
With this formula we can begin thinking about Phase III of a tidal disruption event, and in particular whether it can alter Eq. \ref{DeltaEpsilonU}.

As we have seen in the previous section, once $f\approx f_{\rm c}$, the majority of the star simultaneously ``pancakes'' into a sheet of matter strongly compressed in the vertical direction.  If non-gravitational forces were truly negligible, an idealized, one-dimensional ($\hat{z}$ extent only) star would momentarily possess zero height at $f=f_{\rm c}$, but in reality sufficient compression will create a pressure gradient strong enough to oppose free fall in the z-direction.  The resulting bounce will reverse the vertical free fall and lead to vertical expansion at speeds comparable to $w_{\rm c}$.  The vertical rebound will have a limited impact on $\Delta \epsilon$ because it is effectively decelerated by the tidal potential (for example, if we generously approximate the rebound as elastic due by reflecting $w$ at $f=f_{\rm c}$, the asymptotic free solution velocity $w\to 0$ as $f\to \pi$), but the smaller rebound velocities $\Delta v_{\rm x}, \Delta v_{\rm y}$ in the orbital plane can in principle have more significant effects, as $\Delta \epsilon \sim V_{\rm p}\Delta v$.  In this section we assume the bounce is adiabatic; in particular, we neglect both dissipation in shocks and the thermonuclear energy release from the compression of the stellar core.  For a more thorough discussion of these possibilities see \citet{LuminetPichon89, BrassartLuminet08}.

During Phase III of a TDE, the requirement that central pressure rises to halt the kinetic energy of collapse implies that the star's peak internal specific energy will be
\begin{equation}
U_{\rm c} \sim \beta^2 U_* (\sqrt{1-\beta^{-1}} +1)^2
\end{equation}
where $U_*=GM_*/R_*$.  Assuming a polytropic equation of state $P=K\rho^\gamma$, and furthermore that strong compression in the $\hat{z}$ direction means that the density enhancement will be due to collapse in $\hat{z}$ alone (since the cross-sectional area within the orbital plane, $\approx \pi r_{\rm long} r_{\rm short}$ remains roughly constant), gives a minimum stellar height and maximum stellar density of
\begin{equation}
\frac{z_{\rm min}}{R_*} \sim \frac{\rho_*}{\rho_{\rm c}} \sim \beta^{-2/(\gamma -1)},
\end{equation}
where $\rho_{\rm *}$ is the mean pre-disruption stellar density.  The duration of maximum compression is a steep power of the impact parameter, specifically 
\begin{equation}
\tau_{\rm c} \sim \beta^{-(\gamma+1)/(\gamma-1)}\tau_*,
\end{equation}
with $\tau_*=1/\sqrt{G\rho_*}$.  Although we have only derived these formulas at the order of magnitude level, they have been calibrated over a wide range of $\beta$ by both the affine model \citep{LuminetCarter86} and one-dimensional hydrodynamical simulations \citep{BrassartLuminet08} (hereafter BL08).  Specifically, for $\gamma=5/3$ polytropes, the affine model found $U_{\rm c}\approx 1.2 U_*\beta^2$, $\rho_{\rm c}=1.3\rho_*\beta^3$ and $\tau_{\rm c}=8.5\tau_*\beta^{-4}$, calibrations which were essentially duplicated in BL08.  Likewise, $z_{\rm min}\approx 4.5 \beta^{-3} R_*$ if we assume that the rise in density comes entirely from homologous, vertical stellar collapse (i.e. that the in-plane area of the compressed star is $\sim r_{\rm long} r_{\rm short}$).

If we assume that the pressure-driven bounce acts isotropically (i.e. that shear stresses from viscosity or shocks remain unimportant), then the relevant changes in velocity can be estimated as $\rho_{\rm c}\Delta v_{\rm i}/\tau_{\rm c} \sim \Delta P_{\rm c}/r_{\rm i}$, with $r_{\rm i}$ the physical dimension of the star parallel to which pressure gradients impart $\Delta v_{\rm i}$.  Specifically,
\begin{align}
\Delta v_{\rm z} &\sim \sqrt{U_{\rm c}} \frac{z_{\rm min}}{z_{\rm min}}\notag  \\ 
\Delta \vec{v}_{\rm short}\cdot \hat{V_{\rm c}} &\sim \sqrt{U_{\rm c}} \frac{z_{\rm min}}{r_{\rm short}}\sin (\nu_{\rm c}) \\
\Delta \vec{v}_{\rm long}\cdot \hat{V_{\rm c}} &\sim \sqrt{U_{\rm c}} \frac{z_{\rm min}}{r_{\rm long}}\cos (\nu_{\rm c}), \notag
\end{align}
where we have denoted the direction of the orbital velocity at $f=f_{\rm c}$ as the dimensionless unit vector $\hat{V_{\rm c}}$.  

\begin{figure}
\includegraphics[width=85mm]{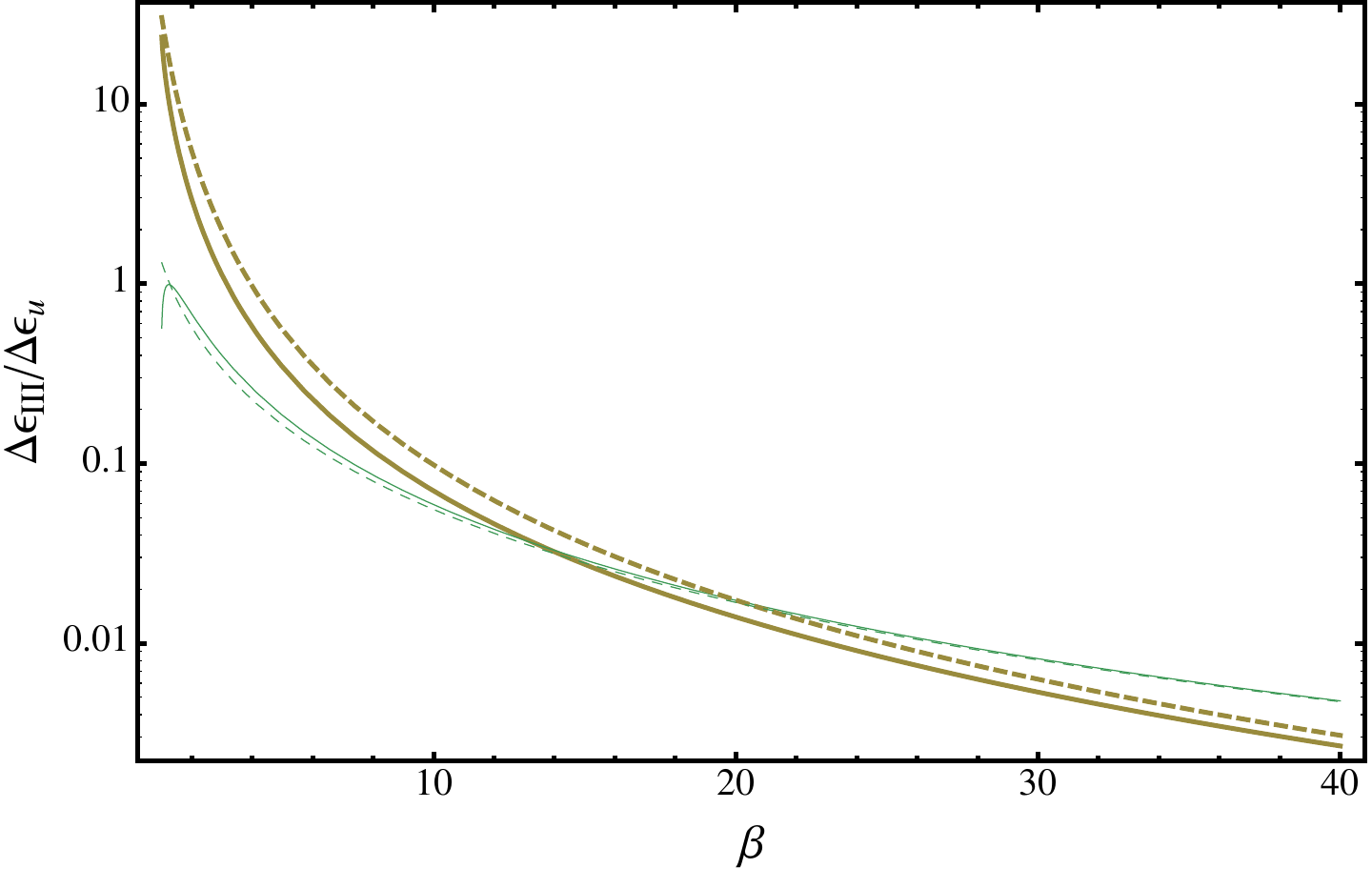}
\caption{Fractional specific energy perturbations during the bounce phase.  We plot perturbations along the short axis of the star as thick yellow curves, and along the short axis as thin green curves.  Exact solutions are solid lines, and the leading order behavior from Eqs. (\ref{deltaEShortApprox}) and (\ref{deltaELongApprox}) are dashed lines.  The bounce represents at most a factor $\approx 2$ correction to specific energy of the stellar debris for $4\gtrsim \beta \gtrsim 3$; above these values, the bounce is negligible.  Below $\beta\approx 3$, a larger correction is possible, but the free solutions become somewhat unreliable.  The high degree of alignment between the stellar bulge and the orbital velocity vector causes perturbations along the long axis to actually dominate those along the short axis above $\beta \approx 15$.}
\label{vEIII}
\end{figure}

This leads to energy perturbations within the orbital plane at bounce of $\Delta \epsilon_{\rm III} \sim V_{\rm c}\Delta v$.  Using our exact formulae for axis lengths and alignment, we plot the results in Fig. \ref{vEIII}, along with the limiting behavior at high $\beta$, which is well-approximated by the Taylor expansions in Appendix A as:
\begin{align}
\Delta \epsilon_{\rm III, short} \sim 31\beta^{-5/2} \Delta\epsilon_{\rm u} \label{deltaEShortApprox} \\ 
\Delta \epsilon_{\rm III, long} \sim 9\beta^{-2}\left(1+\frac{11}{2\beta}\right)^{-1}\Delta\epsilon_{\rm u} . \label{deltaELongApprox} 
\end{align}
Here we have assumed $\gamma=5/3$, and that $z_{\rm min}\approx 5R_*\beta^{-3}$ based on the BL08 calibration.  We have also approximated $V_{\rm c} \approx V_{\rm p}$, which is accurate for high $\beta$ though a mild overestimate at low $\beta$.  Even with this overestimate, we can see from Fig. \ref{vEIII} that only for $\beta \lesssim 3$ (where our model's assumption of tidal free fall begins to break down) can the pressure-driven bounce along the short principal axis of the star provide an order unity enhancement to the total spread in (in-plane) debris energy.  The contribution of the bounce along the longer principal axis remains negligible at all $\beta$.  

Because the z-component of $ \vec{V}_{\rm c} =0$, the spread in kinetic energy of vertical motion at the time of bounce is given by
\begin{equation}
\Delta \epsilon_{\rm III, z} \sim \Delta v_{\rm z}^2 \sim \beta^2 \frac{GM_{\rm *}}{R_*}.
\end{equation}
Interestingly, for $\beta \gtrsim 10$ the total spread in kinetic energy at the time of bounce is dominated by $\Delta\epsilon_{\rm III, z}$, not $\Delta\epsilon_{\rm u}$.  However, the instantaneous vertical kinetic energy at $f=f_{\rm c}$ will disappear as $f \to \pi$.  Even neglecting dissipation of the kinetic energy of vertical free fall into shocks, and assuming a perfectly elastic bounce, the tidal potential (Eq. \ref{selfSimilar}) will efficiently decelerate the vertical motion of the debris during the phase IV rebound and later expansion.  This can be seen by continuing the free solutions past $f=f_{\rm c}$, which corresponds to a reflection of vertical velocity.

Here we have ignored energy release from thermonuclear burning at the time of maximum compression, which in principle could increase $\Delta \epsilon$.  However, past estimates made in the framework of the affine model \citep[Table 8]{LuminetPichon89} found that for the range of $\beta$ values considered ($5 \le \beta \le 20$), the total thermonuclear energy release was less than $U_{\rm c}$, making it unlikely to change the analysis of this section.

Of course, our analysis of energy redistribution in Phase III depends critically on how synchronously the vertical collapse of the star proceeds.  If individual ``columns'' of the star do not collapse in the homologous manner implied by Eq. \ref{selfSimilar}, it is unlikely $z_{\rm min}$ will reach the extreme values predicted by the simple arguments in this section.  Alternatively, if separate columns collapse in a desynchronized way, it is possible that pressure waves from collapsed regions of the star will propagate upstream to uncollapsed regions and cause them to rebound prematurely.  In either scenario, the effective $z_{\rm min}$ will be enhanced, enabling greater coupling of the bounce energy to motions within the orbital plane, and increasing the values of $\Delta\epsilon_{\rm III, short}$ and $\Delta\epsilon_{\rm III, long}$.  Therefore, Eq. \ref{DeltaEpsilonU} should be regarded as a lower bound on $\Delta \epsilon$ - a higher value of $n$ would be favored if  the desynchronization of vertical collapse transfers kinetic energy to in-plane motions more efficiently than in our estimates here.  A similar effect should arise in hydrodynamical simulations of TDEs that lack sufficient vertical resolution to capture the maximum compression of the star (we discuss this further in \S 9).  In the following section, we consider physical sources of desynchronization.

\section{Desynchronization}

The synchronous vertical collapse of a one-dimensional star into a thin, pancake-like sheet only occurs if the initial distribution of vertical velocities is self-similar, i.e. $w_0(z_0)\propto z_0$.  In previous sections we have assumed the trivial self-similarity of $w_0=0$.  Deviations from self-similarity will be seeded at early times by the nonlinear hydrodynamics of actual disruption at the tidal radius $R_{\rm t}$, and also later, as the self-gravity and pressure of the stellar debris perturbs the free solutions for $f<f_{\rm c}$.  In this section we quantify in an approximate way the effect of desynchronization on our idealized earlier conclusions, finding that both the stellar properties during Phase III (important for any shock breakout signal) as well as $\Delta \epsilon$ could be significantly altered.  However, we then argue that past hydrodynamical simulations indicate that desynchronization is likely to be suppressed in physical TDEs, justifying our use of the parabolic free solutions.  Finally, we consider the desynchronization of stellar collapse in three dimensions.

\subsection{Desynchronized Free Solutions}

At $f=f_{\rm t}$, during the transition from Phase I to Phase II, velocity perturbations of order $\sim \sqrt{GM_*/R_*}$ could be imprinted on the free-falling stellar debris.  Normalizing our initial conditions $\{z_0, w_0\}$ in units of $R_*$ and $\sqrt{GM_*/R_*}$, we derive coefficients for the ``perturbed'' (i.e. $w_0 \ne 0$) vertical free solutions to be 
\begin{align}
&E_{\rm p}=-\tilde{z}_0 \sqrt{\beta-1}- \tilde{w}_0\sqrt{\frac{1}{2\beta}}(\beta-2) \\
&F_{\rm p}=\tilde{z}_0+\tilde{w}_0\sqrt{\frac{2}{\beta}}\sqrt{\beta-1}. 
\end{align}
Therefore the true anomaly of a perturbed vertical collapse to $z=0$ is 
\begin{equation}
\tan(f_{\rm c}^\prime)=\frac{\tilde{z}_0+\tilde{w}_0\sqrt{2}\sqrt{\beta-1}/\sqrt{\beta}}{\tilde{z}_0\sqrt{\beta-1}+\tilde{w}_0\sqrt{2}(\beta/2-1)/\sqrt{\beta}}. \label{fcprime}
\end{equation}

We note that both $f_{\rm c}$ and $f_{\rm c}^\prime$ go $\propto \beta^{-1/2}$ in the large $\beta$ limit.  Unless  $w_0 \propto z_0$, the collapse will be non-homologous, with $f_{\rm c}^\prime$ depending on $z_0$.  Modest deviations from homologous initial conditions will desynchronize the collapse, which we illustrate by plotting the desynchronized free solutions for $\beta=2$ and $\beta=10$.  We can see that in both cases, the time at which the free solutions cross the orbital plane becomes strongly desynchronized, which complicates our previously simple treatment of the transition from ``tidal free fall'' to ``pressure-driven bounce'' and also raises the possibility that the vertical kinetic energy of free fall could be effectively isotropized and transferred to motions within the orbital plane, restoring a $\beta$ dependence to $\Delta \epsilon$.

\begin{figure}
\includegraphics[width=85mm]{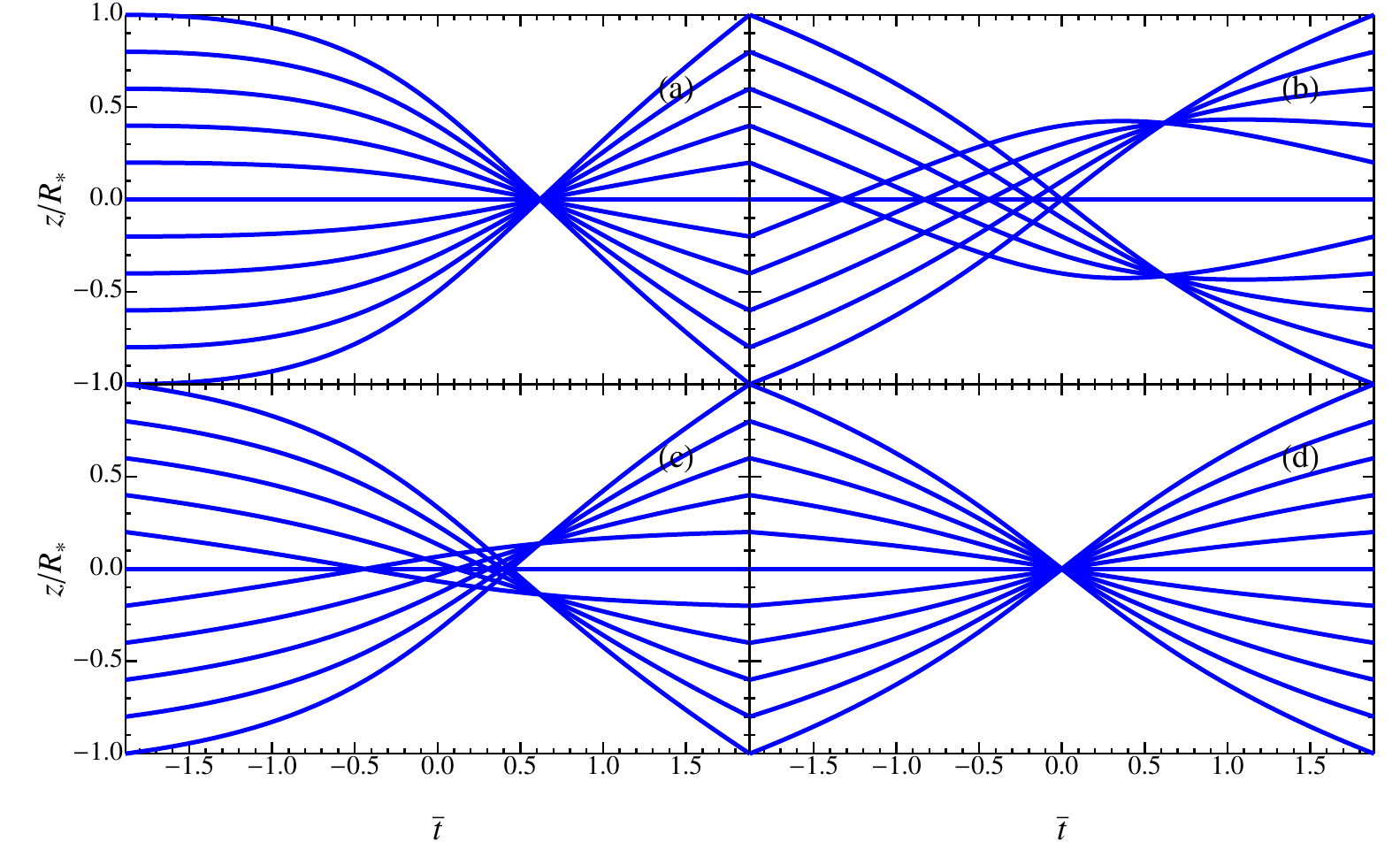}
\caption{Here we plot several sets of free solutions ($\tilde{z}$ versus $\bar{t}$) for $\beta=2$.  In panel a, the initial vertical velocity $w_0=0$.  In panel b, all fluid elements in the star receive initial velocity perturbations $|w_0|=\sqrt{GM_*/R_*}$; in panel c, $|w_0|=\sqrt{GM_*/R_*}/3$.  In panel d, the star receives homologous velocity perturbations $w_0=-\tilde{z}_0\sqrt{GM_*/R_*}$, causing synchronous collapse {\it before} pericenter passage.}
\label{desynch2}
\end{figure}

\begin{figure}
\includegraphics[width=85mm]{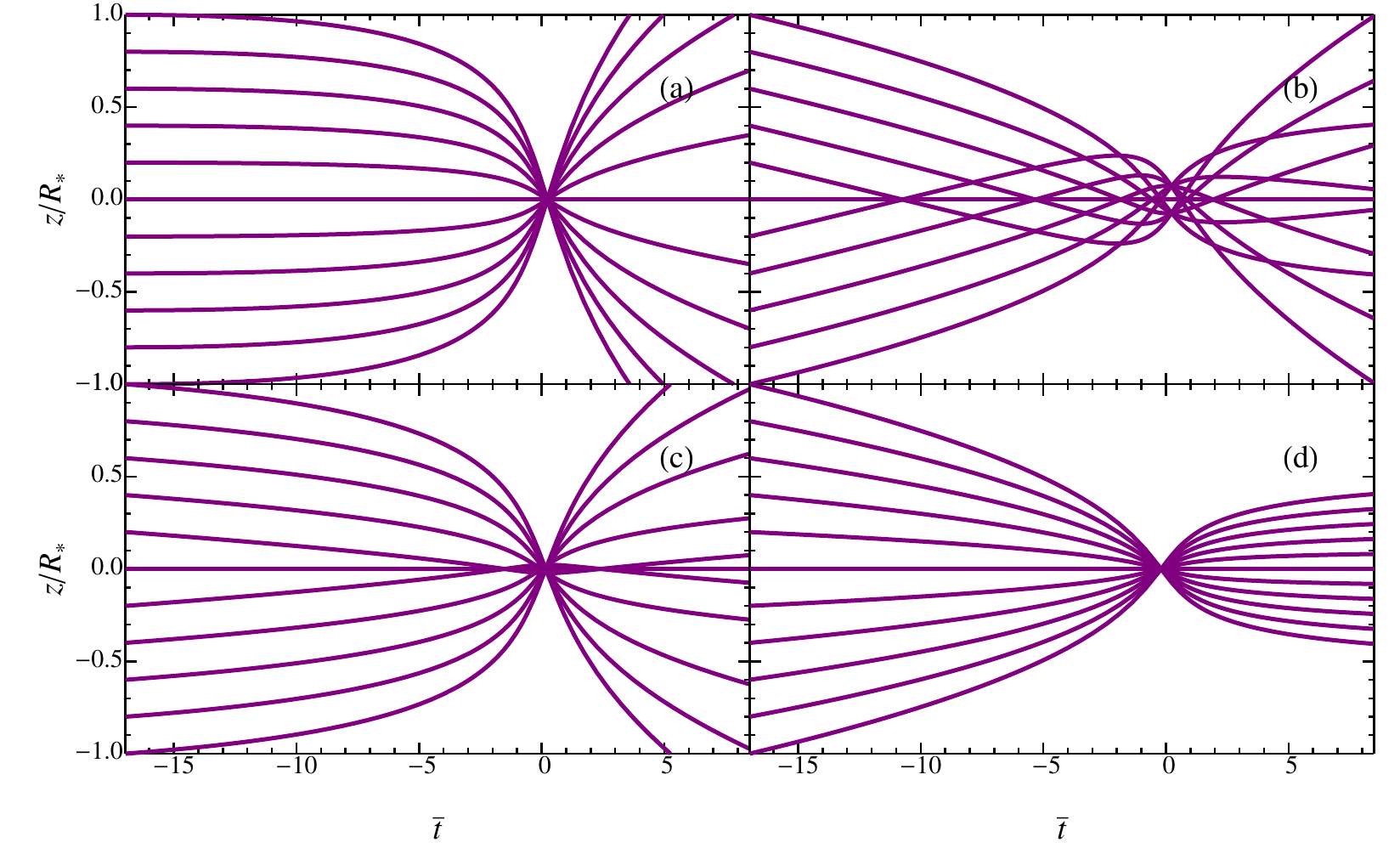}
\caption{The same as Fig. \ref{desynch2}, but for $\beta=10$.  Desynchronization is less severe at higher $\beta$.}
\label{desynch10}
\end{figure}

We can estimate the amount of desynchronization by using trigonometric identities and Eqs. \ref{fc} and \ref{fcprime} to find $\Delta f=f_{\rm c}^\prime - f_{\rm c}$.  Specifically,
\begin{equation}
\tan(\Delta f)=\frac{\tilde{\lambda}_0}{\sqrt{2\beta}+\tilde{\lambda}_0\sqrt{\beta-1}},
\end{equation}
where $\tilde{\lambda}_0=\tilde{w_0}/\tilde{z_0}$.

Interestingly, both Figs. \ref{desynch2} and \ref{desynch10} show that most of the star's desynchronized free solutions have two crossings of the orbital plane, raising the possibility of a double bounce in desynchronized collapse scenarios (something previously seen only due to GR effects, e.g. \citet{LuminetMarck85} - see \S VI).  But is it realistic to expect desynchronized collapse?

\subsection{Validity of Free Solutions in One Dimension}

From the above discussion, it is clear that only modest deviations from self-similarity in the initial velocity perturbations $w_0$ will produce a strongly non-homologous vertical collapse at most realistic $\beta$.  We can quantify the magnitude of the initial velocity perturbations $\tilde{\lambda}_0$ required to significantly desynchronize one-dimensional collapse by making the approximation (valid for small $\Delta f$) that the desynchronization timescale $\Delta t_{\rm 1D}\approx \Delta f \sqrt{R_{\rm p}^3/(GM_{\rm BH})}$.  If we then require $\Delta t_{\rm 1D}<\tau_{\rm c}= \chi_{\rm c} \tau_*$, then for a $\gamma=5/3$ polytrope (with $\chi_{\rm c}\approx 8.5$) and inwardly directed velocity perturbations we find the condition that 
\begin{equation}
|\tilde{\lambda}_0|\lesssim \frac{\sqrt{2\beta}\tan(2\chi_{\rm c} \beta^{-5/2})}{1+\sqrt{\beta-1} \tan(2\chi_{\rm c}\beta^{-5/2})}. \label{spinCondition}
\end{equation}
The factor of $\approx 2$ inside the argument of the tangent comes from the difference between $\tau_*=\sqrt{1/G\rho_*}$ and $\sqrt{R^3/(GM_*)}$.  This condition grows more restrictive as $\beta$ increases, with the right hand side of Eq. \ref{spinCondition} roughly proportional to $\beta^{-2}$.

For one-dimensional stellar collapse, high-resolution hydrodynamical simulations indicate that a highly homologous collapse is physically realized (BL08).  As noted before, this is likely due to a combination of two factors: the partial cancellation of stellar pressure with self-gravity, and also that $a_{\rm t}/a_{\rm g}\approx(R_{\rm t}/R)^3$.  This explanation is supported by past investigations of stellar tidal disruption in the affine ellipsoids approximation: for example, Fig. 4 in CL83 shows the first order cancellation of pressure and self-gravity for early parts of Phase II.  Three results of BL08 further support the validity of the unperturbed free solutions in Phase II of a TDE:
\begin{itemize}
\item The actual collapse of the star is visually homologous during Phase II, as seen by the near-linearity of a vertical velocity versus height plot at different times (BL08, Fig. 2).  Although the figure deviates slightly from homologous collapse at large radii, possibly due to the fact that the low-density outermost regions of the star are disrupted slightly before the higher density inner regions (like the peeling of onion shells), these outer deviations do not appear to affect the key dynamics of the bounce.
\item The maximum central compression $\rho_{\rm min}\sim \beta^{2/(\gamma-1)}\rho_*$ in accordance with the assumption of fully synchronized tidal free fall (BL08, Eq. 43).  We note that this is a geometric proxy for $z_{\rm min}$.
\item The bounce of the collapsing star occurs after pericenter passage (BL08, Table 5).  As shown above, this places a strong constraint on the initial velocity perturbations.  In particular, let us consider a perfectly homologous collapse for the sake of argument, with $\tilde{w}_0=-\tilde{\lambda}_0 \tilde{z}_0$.  Eq. \ref{fcprime} will only be positive (i.e. bounce after pericenter passage) if $\tilde{\lambda}_0<(\sqrt{2}\beta\sqrt{\beta-1})^{-1}$, a rather small perturbation ($\tilde{\lambda} <0.04$ for $\beta=7$, as is relevant here).
\end{itemize}

These numerical results indicate that realistic one-dimensional stars behave during Phase II much like the unperturbed free solutions we presented in \S 3, supporting our earlier assumption that debris energy ``freezes in'' from $f=f_{\rm t}$ down to the bounce, at $f=f_{\rm c}$.  

\subsection{Validity of Free Solutions in Three Dimensions}
The full problem of tidal disruption is three dimensional, and some three dimensional simulations \citep{Laguna+93, Guillochon+09} have indicated that one-dimensional descriptions of the bounce phase \citep{LuminetCarter86} may strongly overestimate the degree of compression.  However, lack of vertical resolution in the three dimensional simulations makes it difficult to interpret the discrepency, and some high resolution simulations \citep{RosswogRR09} do find degrees of compression closer to our analytic expectations in \S 4.  Although the impact of higher dimensional effects on Phase III of a TDE will only be resolved through higher resolution hydrodynamical simulations, we present here a simple analytic argument suggesting that a star made of many columns, each undergoing homologous collapse, should attain $z_{\rm min}$ comparable to one-dimensional predictions.

Three dimensional desynchronization is an important effect that cannot be ignored at high $\beta$: the bounce timescale $\tau_{\rm c}\approx 8.5\tau_*\beta^{-4}$ for $\gamma=5/3$, while the time it takes the bulk of the star to pass across the tidal radius is $\Delta t_{\rm 3D} \approx 1.4 \tau_* (M_*/M_{\rm BH})^{1/3}$.  This implies that for $\beta$ larger than a critical value,
\begin{equation}
\beta_{\rm d}=1.6\left(\frac{M_{\rm BH}}{M_*}\right)^{1/12},
\end{equation}
$\tau_{\rm c} \ll \Delta t_{\rm 3D}$, and the leading edge of the star will collapse and rebound well before the trailing edge.  If we assume that the star is truly in tidal free fall during Phase II, and seed velocity perturbations remain as small in three dimensions as has been indicated in one dimensional simulations, then each column of the star will reach its maximum compression at different $t$ but the same $f_{\rm c}$. 

This $f_{\rm c}$ remains fixed in space, much like a nozzle, as the star passes through it.  An example of this ``tidal nozzle'' has been seen in hydrodynamical simulations of tidal disruptions of white dwarfs; for example, Fig. 6 of \citet{RosswogRR09}.  Even if we assume the maximum compression predicted by one-dimensional models of stellar collapse ($z_{\rm min}\approx 4.5R_*\beta^{-3}$, for $\gamma=5/3$), the sound speed in the stellar midplane, $c_{\rm s, c}$, will remain a small fraction of the stellar orbital velocity.  Specifically $c_{\rm s, c}/V_{\rm p}\approx \beta^{1/2}(M_*/M_{\rm BH})^{1/3}$, indicating that unphysically large $\beta$ values are required for pressure waves from the region of maximum compression to communicate upstream to the Phase II material.  Unless three dimensional effects influence earlier stages of a TDE (by seeding large perturbations during the transition from Phase I to Phase II), it seems unlikely that the star will be prevented from reaching the strong compressions suggested by models of one-dimensional collapse.  Among other things, this highlights the importance of thermonuclear network calculations for high-$\beta$ TDEs \citep{LuminetPichon89}.

With these caveats in mind, we generalize the work of \S III to perturbed in-plane free solutions, i.e. where every fluid element at $f=f_{\rm t}$ has initial positions $\{x_0, y_0\}$ but also initial velocities $\{u_0, v_0\}$.  The in-plane coefficients for the corresponding ``perturbed'' free solutions are

\begin{align}\label{xyuvICs}
A_{\rm p}=&\frac{1}{\beta^2} \Big(-8 \tilde{x}_0 \sqrt{\beta -1} + 2\tilde{y}_0 (\beta^2+2\beta -4)\Big) \\
&+ \frac{2\sqrt{2}}{\beta^{3/2}}\Big (\tilde{u}_0(2-3\beta)+\tilde{v}_0\sqrt{\beta-1}(\beta-2)\Big) \notag \\
B_{\rm p}=&\frac{1}{5\beta^2} \Big(2 \tilde{x}_0 \sqrt{\beta -1} (\beta^3-4\beta^2+8) + \tilde{y}_0 (9\beta^3\\
&-12\beta^2-8\beta+16)\Big) +\frac{1}{5\sqrt{2}\beta^{3/2}}\Big(\tilde{u}_0(\beta^3-8\beta^2\notag \\
&+28\beta-16)+2\tilde{v}_0\sqrt{\beta-1}(3\beta^2-6\beta+8) \Big) \notag \\
C_{\rm p}=&\frac{1}{\beta^2} \Big(\tilde{x}_0 (2\beta^2+\beta-2) - 2 \tilde{y}_0 \sqrt{\beta-1} (\beta^2 -1) \\
&-\sqrt{2\beta} \big(\tilde{u}_0(1-2\beta)\sqrt{\beta-1}+\tilde{v}_0(\beta-1)^2\big )\Big)\notag \\
D_{\rm p}=&-\frac{1}{20\beta^2} \Big(\tilde{x}_0 (\beta - 2) + 2 \tilde{y}_0 \sqrt{\beta -1}\\
&-\sqrt{2\beta}(\tilde{u}_0\sqrt{\beta-1}+\tilde{v}_0)\Big). \notag
\end{align}

\begin{figure}
\includegraphics[width=85mm]{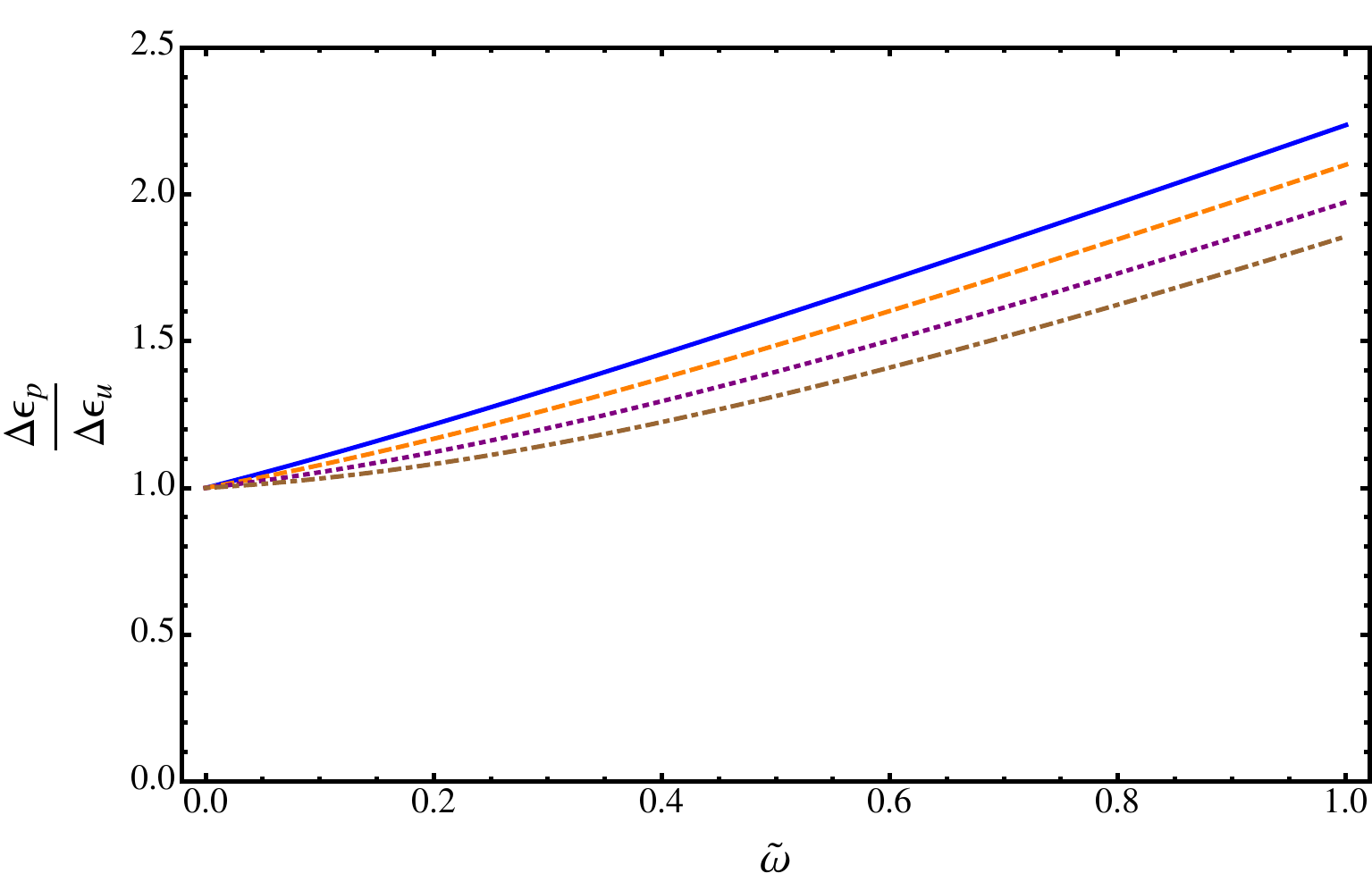}
\caption{The enhancement $\Delta\epsilon_{\rm p}/\Delta\epsilon_{\rm u}$ to the energy spread for initially unperturbed free solutions, when pre-disruption stellar spin (along an axis parallel to orbital angular momentum) is considered.  The energy spread is plotted against pre-disruption spin $\tilde{\omega}$ , where $\tilde{\omega}$ is stellar spin normalized by the breakup spin $\sqrt{GM_*/R_*^3}$.  As in previous plots, the solid blue, dashed orange, dotted purple, and dot-dashed cyan curves represent $\beta=2, \beta=4, \beta=10$, and $\beta=40$, respectively.}
\label{stellarSpin}
\end{figure}

If we now calculate the perturbed specific energy of the free solutions, we find
\begin{align}
\epsilon_p=&\frac{GM_{\rm BH}R_*}{R_{\rm t}^2} \Big(\tilde{x}_0(1-2/\beta)+2\tilde{y}_0\sqrt{\beta^{-1}-\beta^{-2}}\label{epsilonP}  \\
&-\sqrt{2/\beta}(\tilde{u}_0\sqrt{\beta-1}+\tilde{v}_0)\Big ),\notag  
\end{align}
where the initial velocities have been normalized by $\sqrt{GM_*/R_*}$.  Again, there is no leading order $\beta$ dependence in the specific energy, although a calculation of $\Delta\epsilon_{\rm p}$ does not find it completely $\beta$-independent as in Eq. \ref{DeltaEpsilonU}.  Nonetheless, the assumption of tidal free fall during Phase II implies clearly that the frozen-in $\Delta \epsilon$ should be, to leading order, independent of $\beta$.  As a simple test case, we now apply these perturbed free solutions to a uniformly spinning star, with normalized angular velocity $\tilde{\omega}=\omega/\sqrt{GM_*/R_*^3}$ such that $\tilde{\omega}=1$ is approximately the breakup frequency (and spin parallel to orbital angular momentum).  In Eq. \ref{epsilonP}, we relabel $\tilde{x}_0=\tilde{r}_0 \cos\phi_0$, $\tilde{y}_0=\tilde{r}_0 \sin\phi_0$, $\tilde{u}_0=-\tilde{\omega} \tilde{r}_0 \sin\phi_0$,  $\tilde{v}_0=\tilde{\omega} \tilde{r}_0 \cos\phi_0$, and then extremize $\Delta \epsilon$ with respect to $\phi_0$.  Results are plotted in Fig. \ref{stellarSpin}; in general, pre-disruption stellar spin will enhance the energy spread $\Delta \epsilon$ by a small factor, $\lesssim 2$.  Low $\beta$ and high $\tilde{\omega}$ will maximize the energy spread.  

\begin{figure}
\includegraphics[width=85mm]{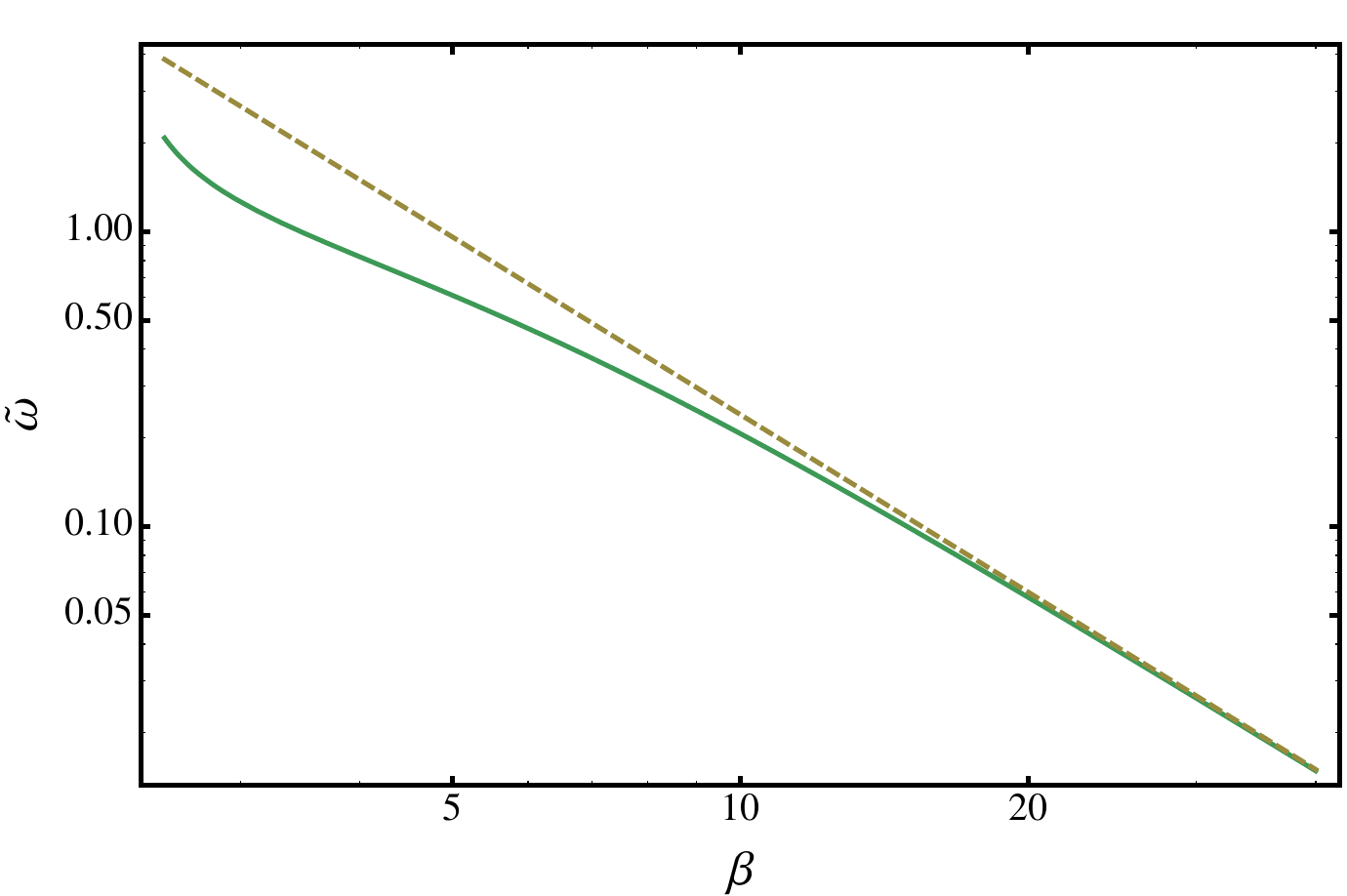}
\caption{The maximum value of normalized stellar spin $\tilde{\omega}$ that will not produce significant 1D desynchronization leading into the phase III bounce.  We plot the exact value calculated from Eq. \ref{spinCondition} as a solid green line, and the asymptotic behavior $\tilde{\omega}\lesssim 2\sqrt{2}\chi_{\rm c}\beta^{-2}$ as a dashed yellow line.  Regions above the curves will experience desynchronization of vertical collapse.}
\label{spinDesynchronization}
\end{figure}

We note here that large stellar spins misaligned with the orbital angular momentum vector could have a much greater impact on $\Delta\epsilon$ by inducing vertical desynchronization.  A thorough investigation of misaligned spin effects is beyond the scope of this work, but as an idealized limiting case we apply Eq. \ref{spinCondition} to approximate, as a function of $\beta$, the maximum stellar spin allowed before the Phase III bounce would be vertically desynchronized.  Specifically, we set $w_0=\omega_0r_0$.  We plot these results in Fig. \ref{spinDesynchronization}, and find that combinations of high $\beta$ and relatively rapid values of stellar spin are needed to strongly desynchronize the vertical collapse.

Fig. \ref{stellarSpin} can be taken as representative of the effects of both primordial stellar spin, and the angular momentum imparted during tidal spin-up of the star prior to its full disruption, during the transition between phases I and II of a TDE.  Tidal spin-up is unlikely to produce misaligned spin, however, so the more dramatic type of desynchronization suggested by Fig. \ref{spinDesynchronization} can only come from the star's original, pre-disruption spin.

For one dimensional stellar collapse, the frozen-in $\Delta\epsilon$ will dominate the post-bounce $\Delta\epsilon$ for all $\beta$.  For three dimensional collapse, the numerical literature is less clear, but we have argued here that three-dimensional effects are unlikely to strongly redistribute energy to in-plane motion, with the possible exception of when sufficiently rapid stellar spin is misaligned with the orbital plane, or perhaps when a similar misalignment between SMBH spin and orbital angular momentum exists.

\section{General Relativistic Corrections}

The results of all prior sections have assumed purely Newtonian gravity; however, tidal disruption occurs at an orbital distance $R_{\rm t}\lesssim 50R_{\rm g}$, where the gravitational radius $R_{\rm g}=GM_{\rm BH}/c^2$.  At these small separations, ballistic motion follows the geodesics of the Schwarzschild or Kerr metric rather than free-fall trajectories in Newtonian gravity: general relativity (GR) is important.  A fully relativistic analysis of the problem of tidal disruption is beyond the scope of this paper, although it has been treated in the past in the case of the affine model \citep{LuminetMarck85}, and in one-dimensional hydrodynamical simulations \citep{BrassartLuminet10}.  If we treat the internal dynamics of the star in a Newtonian way (i.e. assume tangentially flat space-time in the small region occupied by the star), then Eq. \ref{selfSimilar}'s description of vertical collapse will be modified, to become
\begin{equation}
\ddot{z}_{\rm GR}=\frac{\partial\Psi}{\partial z}\left(1+3\frac{L^2}{R^2}\right),
\end{equation}
where $L$ is the orbital angular momentum of the star, $R$ is the orbital radius of the star (both in geometrized units) and we have limited ourselves to non-spinning black holes.  The qualitative results of both \citet{LuminetMarck85, BrassartLuminet10} were that the increased strength of the GR tidal field (relative to Newtonian gravity) can actually result in multiple vertical collapses, each followed by separate bounces which are reversed by the relativistically enhanced tidal field.  For all but the most deeply-plunging TDEs ($\beta \gtrsim 30$), the maximum compression is obtained on the first vertical collapse and is similar to the Newtonian $z_{\rm min}$ \citep[Fig. 10]{LuminetMarck85}.  Therefore, even though the formation of multiple outgoing shockwaves could be an important outcome of relativistic compression, the first-order spread in debris energy is unlikely to be affected by multiple compressions for $\beta \lesssim 30$.

A separate relativistic effect concerns modifications to the pre-bounce spread in debris energy, $\Delta \epsilon$.  Eq. \ref{CorrectedEnergy} was derived by Taylor expanding the Newtonian gravitational potential about the star's position when it crossed into the tidal sphere, then subtracting the zeroth-order component.  We will now repeat that procedure for a post-Newtonian (PN) effective potential which incorporates leading-order GR effects for non-spinning, Schwarzschild black holes.  Specifically, we use the 1PN harmonic coordinate Lagrangian presented in \citet[Eq. 174]{Blanchet06}:
\begin{align}
&\mathcal L^{\rm harm}=\frac{Gm_1m_2}{2r_{12}}+\frac{m_1v_1^2}{2}+\frac{1}{c^2} \Big( -\frac{G^2m_1^2M_2}{2r_{12}^2} + \frac{m_1v_1^4}{8} \\
&+ \frac{Gm_1m_2}{r_{12}} \Big( -\frac{1}{4}(\vec{n}_{12}\cdot \vec{v}_1)(\vec{n}_{12}\cdot \vec{v}_2) + \frac{3}{2}\vec{v}_1^2-\frac{7}{4}(\vec{v}_1 \cdot \vec{v}_2)\Big) \Big) \notag
\end{align}
We define the effective potential as $\Phi_{eff}=K-\mathcal L^{\rm harm}$, where $K$ represents the kinetic energy component of the Lagrangian, i.e. those terms which depend only on velocities.  This equation was derived for arbitrary mass-ratio systems, but here we identify the star as particle 1, the SMBH as particle 2, and have dropped all terms proportional to $v_2$ or $m_1/m_2$.  The Taylor expansion of $\Phi_{eff}$ up to first PN order, around $R=R_{\rm t}$, is given by

\begin{equation}
\Delta \epsilon_{\rm GR} = \frac{GM_{\rm BH}R_*}{R_{\rm t}^2}\left(1+\frac{3V^2}{2c^2}-\frac{R_{\rm g}}{R_{\rm t}} \right),
\end{equation}
where $V$ is the star's orbital velocity at the tidal radius.  

From this equation, it is clear that GR corrections to the Newtonian potential will only matter for large, $M_{\rm BH}>10^7 M_{\odot}$ SMBHs, with tidal radii close to or within the ISCO.  However, all TDEs due to such black holes, or even more massive ones \citep{Kesden11}, will have debris energy spreads modified by GR around the $\sim 2$ level (although we caution that our PN approximation breaks down for tidal radii approaching the ISCO).  In this discussion we have neglected spin effects, but they may also play an important role for the subset of TDEs with $R_{\rm p}\lesssim R_{\rm ISCO}$.  During completion of this paper, a more precise formalism was presented for estimating the GR corrections described in this section \citep{Kesden12}.  The results indicate generally small (factors $\lesssim 3$) GR corrections which are maximized when the spread in energy of a spherical star is calculated near the ISCO, in qualitative agreement with our findings.  We note however that \citet{Kesden12} uses the older, less accurate approach to treating ``frozen-in'' debris energy, i.e. evaluating the spread in energy at $R_{\rm p}$ rather than $R_{\rm t}$.  The primary difference between our approaches is that if debris energy freezes in at the tidal radius, GR corrections to $\Delta\epsilon$ can only reach large, $\sim 2$ levels for $M_{\rm BH} \gtrsim 10^{7.5}M_{\odot}$.  TDEs around low-mass SMBHs will see negligible GR corrections to $\Delta\epsilon$, even if $\beta$ is large.

\section{Gravitational Waves}
The parabolic motion of a star past orbital pericenter will produce low-frequency gravitational waves due to time variation in the quadrupole moment of the star-SMBH system.  For pericenters $R_{\rm p}<R_{\rm t}$, past work has indicated that such a signal could be marginally detectable with a LISA-like instrument \citep{Kobayashi+04}.  Analogous work has focused on the inspiral of a white dwarf (WD) into an intermediate-mass black hole, where a similar signal could be generated by a violent disruption \citep{RosswogRR09, Haas+12}, or a longer-lived GW signal could be accompanied by electromagnetic transients due to inspiral and stable mass transfer \citep{Zalamea+12}.  An alternate, internal source of GWs in TDEs comes from time variation of the star's own quadrupole moment during the Phase III vertical rebound, which was estimated in the past to generate gravitational waves with strain $h \propto \beta^3$ \citep{Guillochon+09}.

In this section, we present more detailed estimates of the ``internal'' GWs due to stellar pancaking and rebound, a process analogous to GW generation during core-collapse supernovae (CCSNe).  As we shall demonstrate, TDE GWs are weakened relative to those in CCSNe due to lower collapse velocities and bounce accelerations, but are increased due to the large degree of stellar asymmetry, and perhaps also by the correspondingly long lever arm of collapse in the quadrupole moment tensor.

Specifically, we consider GW emission at the moment of maximum stellar compression, which we for now take to be synchronized throughout the star (but which will actually occur at different times for each point in the star, as seen in \S 5).  The two polarization components of a GW signal, $h_+$ and $h_{\times}$, can be read off of the transverse traceless GW strain
\begin{equation}
{h}_{\rm ij}^{\rm TT}=\frac{2G}{dc^4}\ddot{J}_{\rm ij}^{\rm TT},
\end{equation}
where $d$ is the distance from the observer to the source, $J_{ij}=I_{ij}-\frac{1}{3}\delta_{ij}\delta^{kl}I_{kl}$ is the reduced quadrupole moment tensor, $J_{ij}^{\rm TT}$ is a projection of $J_{ij}$, and $I_{kl}$ is the standard quadrupole moment tensor:
\begin{align}
&\ddot{I}_{kl}=\int {\rm d}^3\vec{r}\rho \label{quadMoment}\times  \\
&\begin{pmatrix}
  2\dot{x}^2+2x\ddot{x} & 2\dot{x}\dot{y}+\ddot{x}y+x\ddot{y} & 2\dot{x}\dot{z}+\ddot{x}z+x\ddot{z} \\
  2\dot{x}\dot{y}+\ddot{x}y+x\ddot{y} & 2\dot{y}^2+2y\ddot{y} & 2\dot{y}\dot{z}+\ddot{y}z+y\ddot{z} \\
  2\dot{x}\dot{z}+\ddot{x}z+x\ddot{z} & 2\dot{y}\dot{z}+\ddot{y}z+y\ddot{z} & 2\dot{z}^2+2z\ddot{z} \notag
 \end{pmatrix}. 
\end{align}
 To order of magnitude in the limit of fully synchronous vertical collapse, and neglecting the (weak) $\beta$ dependence of all $x$ and $y$ terms, we then have
\begin{align}
&\frac{\ddot{I}_{kl}}{M_*R_*^2\tau_*^{-2}}\sim \int {\rm d}^3\vec{r} 
\begin{pmatrix}
  \beta^0\ & \beta^0 & \beta^5 \\
  \beta^0 & \beta^0 & \beta^5 \\
  \beta^5 & \beta^5 & \beta^2
 \end{pmatrix}, \notag
\end{align}
where we have taken $\gamma=5/3$ (as we will for the remainder of this section) and approximated $\ddot{z}\sim\dot{z}/\tau_{\rm c}\sim \beta^5GM_*/R_*^2$.

If only diagonal terms are considered, the second derivative of the quadrupole tensor will be $\propto \beta^2$.  However, the extremely steep $\beta^5$ dependence of the off-diagonal $\ddot{I}_{\rm xz}$ and $\ddot{I}_{\rm yz}$ terms indicates that viewing angles not closely aligned with $\hat{z}$ could in principle observe copious GW production.  We note that physically, $\ddot{I}_{\rm xz}\sim 10\ddot{I}_{\rm yz}$ because for the large $\beta$ where GW emission is relevant, $\hat{y}$ will be aligned with the star's longest principal axis.

However, two degrees of symmetry present in this problem will substantially reduce the magnitudes of $\ddot{I}_{xz}$ and $\ddot{I}_{yz}$.  The free solutions indicate that to lowest order, tidally free-falling bodies should possess reflection symmetry about their in-plane principal axes.  There is also an additional symmetry of reflection about the orbital plane.  If these symmetries are exact, the off-diagonal terms in Eq. \ref{quadMoment} will integrate to 0.  The in-plane symmetries are broken when $R_*/R\sim 1$, i.e. for deeply plunging disruptions around low-mass SMBHs.  The orbital plane reflection symmetry is more robust, and likely can only be broken by misalignment between the orbital plane and SMBH or stellar spin, which is beyond the scope of this paper.  For the remainder of this section, we treat GW emission from off-diagonal terms as speculative, but the large magnitude of these terms in the integrand should motivate future work on disruptions of spinning stars, or TDES around spinning SMBHs.

For $\beta < \beta_{\rm d}$, three-dimensional desynchronization is unimportant and the star collapses almost simultaneously, emitting GWs with a peak frequency of $\approx 1/\tau_{\rm c}$, which, using the calibration of $\tau_{\rm c}\approx 8.5\beta^{-4} \tau_*$ from the affine model and one-dimensional hydro simulations (\S 4), gives
\begin{equation}
f_{\rm GW}\approx 15~{\rm Hz}~ \left( \frac{\beta}{25}\right)^4 m_*^{1/2}r_*^{-3/2},
\end{equation}
where we have normalized $m_*=M_*/M_{\odot}$ and $r_*=R_*/R_{\odot}$.  Low mass stars have an easier time achieving high frequencies; if we use the relation $R_* \propto M_*^{0.8}$ for main sequence stars with $M_* \le M_{\odot}$, we find $f_{\rm GW}\approx 10~{\rm Hz}$ at $\beta=15$, for $M_*=0.1M_{\odot}$. 

These frequencies are located on the far edge of the Advanced LIGO band, although with steep $\beta$ dependence.  Because the three-dimensional desynchronization discussed in \S 5 results in the leading edge of the star collapsing before the trailing edge (which lags by a time $\Delta t_{\rm 3D}$),  the GW signal will be smeared out over a range of frequencies between $1/\Delta t_{\rm 3D}$ and $1/\tau_{\rm c}$ when $\beta>\beta_{\rm d}$.  This smears out the gravitational wave emission by a factor $\Delta t_{\rm 3D}/\tau_{\rm c}$, giving us the strain estimates
\begin{align}
{h}_{+} \approx 1\times 10^{-25} m_*^2 r_*^{-1}  d_{10}^{-1}
\begin{cases}
\left( \frac{\beta}{\beta_{\rm d}} \right)^{2}, &\beta \lesssim \beta_{\rm d} \\
\left(\frac{\beta}{\beta_{\rm d}} \right)^{-2}M_6^{1/6}, &\beta \gtrsim \beta_{\rm d}
\end{cases} 
\\
{h}_{\times} \approx 1\times 10^{-23} ~\Xi m_*^2 r_*^{-1}  d_{10}^{-1}
\begin{cases}
\left( \frac{\beta}{\beta_{\rm d}} \right)^{5}, &\beta \lesssim \beta_{\rm d} \\
\left(\frac{\beta}{\beta_{\rm d}} \right) M_6^{5/12}, &\beta \gtrsim \beta_{\rm d}
\end{cases} 
\end{align}
Here $d_{10}$ is distance to the source normalized to 10 megaparsecs, and for clarity (i.e. to separate diagonal and off-diagonal components of $\ddot{J}_{ij}$ into different polarization states) we have assumed a line of sight along the $y$ axis so that $dh_+=G(\ddot{J}_{xx}-\ddot{J}_{zz})/c^4$, and $dh_{\times}=2G\ddot{J}_{xz}/c^4$.  As in \S 5, $\beta_{\rm d} \approx 6$ is the critical $\beta$ value above which TDEs experience significant three dimensional desynchronization.  We have defined a parameter, $\Xi$ (which is most likely $\ll 1$), to parametrize the unknown degree of reflection asymmetry in Phase III of a TDE.  

The prospects for high frequency GW observation of TDEs involving main sequence stars appear dim, unless $\Xi \gtrsim 0.1$.  If we limit ourselves to GWs from $\ddot{I}_{zz}$, then disruptions of abundant low mass stars have an easier time falling within the Advanced LIGO band, but produce too little strain to be detected; disruptions of solar-type stars produce a barely detectable strain at $d\sim 1~{\rm Mpc}$, but will lie outside the Advanced LIGO band for $\beta \lesssim 25$.

Tidal disruptions of white dwarfs by intermediate mass black holes appear more promising: a WD of mass $1~M_{\odot}$ and radius $6\times10^6~{\rm m}$ disrupted at $\beta=5$ by a $10^4 M_{\odot}$ IMBH will produce $h_{+}\approx 6\times 10^{-24}$ from a distance of $20~{\rm Mpc}$ (neglecting all strain from off-diagonal terms in $\ddot{J}_{ij}$).  The peak emission frequency $f_{\rm GW}\approx 60~{\rm Hz}$, but the existence of IMBHs is sufficiently uncertain that we do not attempt a rate estimate.  We note that the level of emission seen at $\sim 50~{\rm Hz}$ frequencies in full numerical relativity simulations of WD-IMBH disruptions \citep{Haas+12} was approximately $\times 10^3$ smaller than our prediction.  This may not surprising, as the maximum density enhancement seen in these simulations is $\lesssim 10$, not $\sim 200$ as predicted by one-dimensional models ($\gamma=5/3, \beta\approx 10$).  It is likely that this discrepency is at least partially due to insufficient vertical resolution: after disruption, the smallest grid cell in these simulations is $\approx R_{\rm WD}/40$.  Furthermore, Newtonian and pseudo-Newtonian simulations of $\beta \approx 10$ WD TDEs \citep{RosswogRR09} find maximum degrees of vertical compression $\sim 100$, in approximate agreement with the arguments presented in \S 4 and 5.

\section{Observational Implications}
In this section, we discuss the observational implications of a revised $\Delta\epsilon$ for optical transient searches.  In general, we predict longer decay times but lower initial mass fallback rates than prior works \citep{StrubbeQuataert09}.

Using the introduced index $n$, we rederive here a number of important consequences of Eq. \ref{ParamEnergy} in a parametrized way.  The fallback time for the most tightly bound debris is
\begin{equation}
t_{\rm fall}=3.5\times 10^6 ~{\rm sec}~k^{-3/2}\beta^{-3n/2}M_6^{1/2}m_*^{-1}r_*^{3/2}, \label{tFall}
\end{equation}
where we have used the normalizations $M_6=M_{\rm BH}/(10^6M_{\odot})$, $m_*=M_*/M_{\odot}$, and $r_*=R_*/R_{\odot}$.  The rate of mass fallback evolves with time $t$ as 
\begin{equation}
\dot{M}_{\rm fall}\approx \frac{M_*}{3t_{\rm fall}}\left(\frac{t}{t_{\rm fall}}\right)^{-5/3},
\end{equation}
and is initially super-Eddington for disruptions of solar-type stars by SMBHs with $M_{\rm BH} \lesssim 10^{7.5}M_{\odot}$.  Assuming a radiative efficiency $0<\eta<1$, the peak (i.e. time of first pericenter return) mass fallback rate is given by 
\begin{equation}
\frac{\dot{M}_{\rm peak}}{\dot{M}_{\rm Edd}}\approx 133~ \eta_{-1}k^{3/2}\beta^{3n/2}M_6^{-3/2}m_*^2 r_*^{-3/2}, \label{MEdd}
\end{equation}
implying that the maximum black hole mass that can undergo a phase of super-Eddington accretion is given by
\begin{equation}
M_{\rm BH, max}=2.6\times10^7~M_{\odot}~\eta_{-1}^{2/3}k\beta^nm_*^{4/3}r_*^{-1}. \label{MaxMass}
\end{equation}
The mass fallback rate becomes sub-Eddington at a time 
\begin{equation}
t_{\rm Edd}=6.6\times10^7 ~{\rm sec}~ \eta_{-1}^{3/5}k^{-3/5}\beta^{-3n/5}M_6^{-2/5} m_*^{1/5}r_*^{3/5}, \label{tEdd}
\end{equation}
although if $t_{\rm Edd}<t_{\rm fall}$ there is no super-Eddington accretion phase.  Here we have set $\eta_{-1}=\eta/.1$.  Considering the dependences of Eqs. \ref{tFall}, \ref{MEdd}, \ref{MaxMass}, and \ref{tEdd} on $n$ we see that adopting Eq. \ref{CorrectedEnergy} can have dramatic effects on TDEs from stars on deeply plunging ($\beta > 3$) orbits.  In particular, using the correct values of $\Delta \epsilon$ reduces the peak mass fallback rate, decreasing the maximum SMBH mass that can produce a super-Eddington accretion phase.  On the other hand, for TDEs with super-Eddington accretion, the duration of the super-Eddington phase is extended for (realistic) low-$n$ values.

\begin{figure}
\includegraphics[width=85mm]{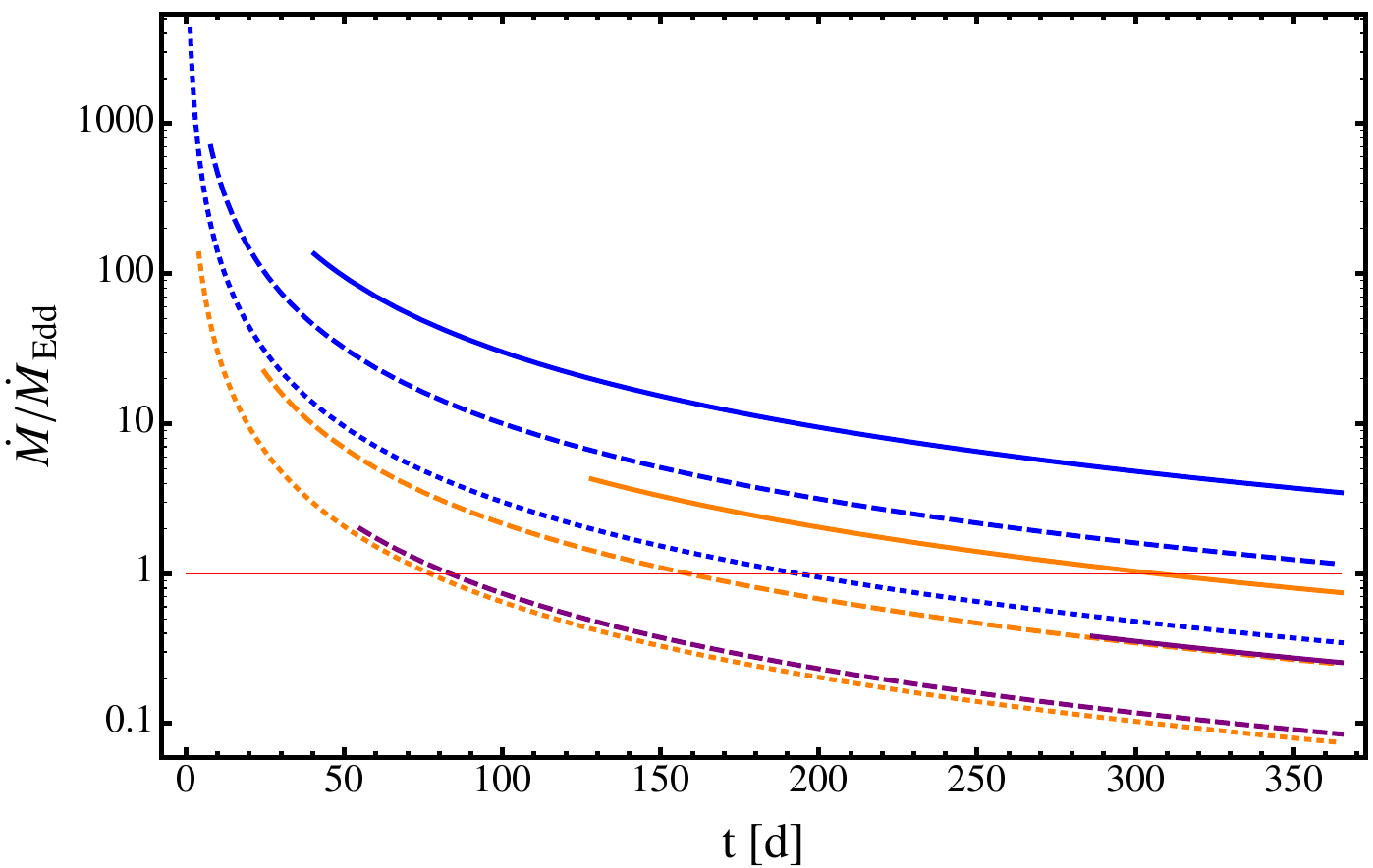}
\caption{Mass fallback curves (normalized by the Eddington fallback rate) versus time since disruption in days, for a variety of TDEs.  SMBH masses of $10^6M_{\odot}$, $10^7M_{\odot}$, and $5\times 10^7 M_{\odot}$ are plotted as blue, orange, and purple curves, respectively.  Likewise, $\beta$ values of 1, 3, and 10 are plotted as solid, dashed, and dotted curves assuming $n=2$, the power law index defined by $\Delta\epsilon\propto \beta^n$ - see Eq. \ref{ParamEnergy}.  If $n=0$, the solid curves represent all $\beta$ values.  Here we consider solar-type stars, and for simplicity set $k=1$ and $\eta=0.1$.}
\label{MFallback}
\end{figure}

We plot the effect of $n$ on the mass fallback rate in Fig. \ref{MFallback}.  In previous literature ($n=2$) a wide variety of mass fallback curves were possible, with high peaks and fast decay times accompanying large $\beta$ values.  If $n=0$, however, the mass fallback rate is generally independent of $\beta$.  Under simplifying assumptions about the relationship of disk luminosity to $\dot{M}$ (often but inaccurately taken as $L\propto \dot{M}$; for complications see \citet{Lodato+09, LodatoRossi11}), the fallback timescale can be inferred by sufficiently long lightcurve observations.  Alternatively, in the future it may be possible to measure $t_{\rm fall}$ directly, by measuring the delay between the onset of accretion and a prompt signal accompanying stellar disruption (either X-ray shock breakout or GWs).  In either case, the $\beta$ independence of $t_{\rm fall}$ will simplify parameter extraction, in particular measurement of $M_{\rm BH}$.

Adopting $n=0$ will also alter the distribution of $\beta$ in the TDEs detected by individual wide-field surveys, $\dot{N}_{\rm det}(\beta)$.  This is a quantity distinct from the distribution of the intrinsic TDE rate, $\dot{N}_{\rm TDE}(\beta)$, which scales as $\dot{N}_{\rm TDE} \propto \beta^{-1}$ for two-body relaxation in the ``pinhole'' regime \citep{BrassartLuminet08}.  Alternatively, if the dominant source of loss cone fueling is two-body relaxation in the ``diffusion'' regime, almost all TDEs will have $\beta=1$; however, since most SMBHs are supplied with stars coming from the boundary between these regimes we will consider the ``pinhole'' regime for the remainder of this section (as it is the relaxational regime with nontrivial $\beta$ dependence).

If we first consider UV or soft X-ray surveys sensitive to the peak frequencies of TDE disk emission, then the peak luminosity $L \propto \dot{M}_{\rm peak} \propto \beta^{3n/2}$, implying a survey horizon $r_{\rm hor} \propto \beta^{3n/4}$ and a detection rate $\dot{N}_{det} \propto \beta^{(9n-4)/4}$.  Optical detections of TDE disks will not be sensitive to the event's bolometric luminosity but rather to emission on the Rayleigh-Jeans tail, for which $L \propto \dot{M}^{1/4}$ and $\dot{N}_{\rm det} \propto \beta^{(9n-16)/16}$ \citep{LodatoRossi11}.  However, both of these scaling relations for $\dot{N}_{\rm det}$ assume a purely flux-limited survey; in the old picture of TDE energy spread ($n=2$), high-$\beta$ events would be favored by their high flux but disfavored by their shorter timescales of peak emission.  If the timescale of peak emission, $t_{\rm fall}$, is less than the survey cadence, $t_{\rm cad}$, then the probability of detection will be approximately reduced by the factor $t_{\rm fall}/t_{\rm cad}$.  This gives $\dot{N}_{\rm det}\propto \beta^{(3n-4)/4}$ and $\dot{N}_{\rm det} \propto \beta^{(-15n-16)/16}$ for X-ray and optical disk emission, respectively.

Although the details remain uncertain, several recent papers \citep{StrubbeQuataert09, LodatoRossi11} have predicted that super-Eddington, radiation-driven outflows may dominate early emission from TDE accretion disks, particularly at long wavelengths.  Using a simple blackbody model with peak frequency $\nu_{\rm bb}$ \citep{LodatoRossi11}, which predicts a peak luminosity $L \propto \beta^{97/24}$ at frequencies $\nu<\nu_{\rm bb}$, or $L\propto \beta^{2/5}$ for $\nu> \nu_{\rm bb}$, we can repeat the above calculations.  The $\beta$ dependence of all different scenarios are presented in Table 1.  With $n=0$, all observational strategies are biased in favor of low $\beta$ detections, including many strategies that once favored high $\beta$ TDEs.

\begin{table}
\centering
\begin{tabular}{ r || r | r | r }
  Scenario & $n=2~(\frac{t_{\rm fall}}{t_{\rm cad}}>1)$ & $n=2~(\frac{t_{\rm fall}}{t_{\rm cad}}<1)$ & $n=0$  \\
  \hline                        
  Disk, $\nu\approx \nu_{\rm bb}$ & 7/2 & 1/2 & -1  \\
  Disk, $\nu< \nu_{\rm bb}$ & 1/8 & -23/8 & -1  \\
  SE, $\nu\approx \nu_{\rm bb}$ & -2/5 & -17/5 & -7/10  \\
  SE, $\nu< \nu_{\rm bb}$ & 81/16 & 33/16 & -33/16  \\
          \end{tabular}
\label{PNValidation}
\caption{Here we display the scaling exponents $s$ for the $\beta$ dependence of the TDE detection rate, $\dot{N}_{\rm det}(\beta) \propto \beta^{s}$.  In the first column we describe the frequency and source of emission (SE indicates a super-Eddington outflow); in the second we give $s$ for the standard scenario $n=2$ and a high-cadence survey; in the third we again consider $s$ in the standard scenario, but for a slow-cadence survey; in the final column we give $s$ for our revised $\Delta\epsilon$, with $n=0$, at any cadence.  In this table $\nu_{\rm bb}$ refers to the ``blackbody'' frequency of peak emission (the super-Eddington outflow is assumed to have a thermal spectrum; the disk emission is better modeled as a multicolor blackbody).}
\end{table}


\section{Discussion}
In this paper, we have analyzed the tidal disruption and free fall of a star in the context of ``free solutions'' to the Hills equations of the parabolic restricted three-body problem.  The important conclusions of this work are the following:

\begin{enumerate}
\item During the tidal disruption of a star, debris energy freezes in at $R=R_{\rm t}$, not $R=R_{\rm p}$.
\item Consequently, the spread in debris energy is smaller than in past analytic predictions.  This will result in flares with longer fallback times, i.e. ones that decay more slowly but have smaller initial fallback rates.  Fewer TDEs will drive powerful super-Eddington outflows than has been predicted in the past.
\item The spread in debris energy is generally dominated by the freeze-in energy, although redistribution of the kinetic energy of vertical collapse to in-plane motions may result in slight variation in $\Delta\epsilon$ for low $\beta (\lesssim 5)$.  Rapidly spinning stars may see a stronger version of this effect at high $\beta$ if their spins are misaligned with the orbital angular momentum vector.
\item The leading order GR corrections to the frozen-out value of $\Delta \epsilon$ are small, and generally negligible unless $R_{\rm p} \lesssim 6R_{\rm g}$.  For such TDEs (i.e. all TDEs for SMBH masses above $10^{7.5}M_{\odot}$) we have derived for the first time the 1PN modifications to $\Delta \epsilon$.  
\item The free solution model we have introduced is, in the limit of spherical and static initial conditions, an approximate simplification of the affine model .  However, the ability to include a range of nonspherical or dynamic initial conditions gives it a degree of flexibility not present in the affine model.  Furthermore, the deformations to an initially spherical body in tidal free fall are not ellipsoids, as is assumed by the affine model.
\item Gravitational waves are generated from variation in the internal quadrupole moment of a tidally disrupting star, reaching peak amplitude at the moment of maximum compression and bounce.  GW emission is likely dominated by a single term in the quadrupole moment tensor, $\ddot{I}_{zz}\propto \beta^2$.  For main sequence stars, these are unlikely to be detectable by ground-based GW interferometers.  GWs from the tidal disruption of WDs by IMBHs are more promising targets, and for modest $\beta$ values ($\sim 5$) could be detectable to tens of megaparsecs.
\item Gravitational wave emission from nondiagonal terms in the star's quadrupole moment could alter the previous conclusion, since $\ddot{I}_{xz}, \ddot{I}_{yz} \propto \beta^5$.  However, in order for these terms to possess a nonzero prefactor, the reflection symmetry of the TDE about the orbital plane must be broken by either SMBH or stellar spin.  Whether this can be done without desynchronizing the vertical collapse and weakening the $\beta$ dependence of $\ddot{I}_{xz}, \ddot{I}_{yz}$ is unclear.
\end{enumerate}

The existing hydrodynamical literature did not until very recently support the first three of these conclusions.  With a few exceptions, prior work has focused mainly on the common $\beta=1$ events, for which Eqs. \ref{UncorrectedEnergy} and \ref{CorrectedEnergy} are identical.  In \citet{Laguna+93}, the authors conducted SPH simulations in a static Schwarzschild background geometry for $\beta=\{1, 5, 10\}$.  They found that the fallback time scaled approximately as $t_{\rm fall} \propto \beta^{-1.5}$ and the velocity of unbound ejecta $v_{\rm ej} \propto \beta^{0.5}$, both of which imply the intermediate value of $n=1$, for the power-law index defined in Eq. \ref{ParamEnergy} as $\Delta\epsilon \propto \beta^n$.  However, the limited resolution of their simulations (7000 SPH particles) makes it unclear whether they would have possessed the midplane resolution to resolve the phase of maximum compression; indeed, they find that $\rho_{\rm c} \propto \beta^{1.5-2}$, a scaling well below analytic predictions as well as the higher-resolution one-dimensional simulations of BL08.  

It is also possible that our approximations have underestimated the efficiency with which the bounce phase can redistribute the energy of vertical collapse to motions within the orbital plane, perhaps due to neglect of GR effects during phase III.  Alternatively, the simulations of \citet{Laguna+93} may be altered by entropy production due to artificial viscosity, as suggested by the authors.

More recent hydrodynamical simulations of high-$\beta$ disruptions did not publish details on the spread of debris energy \citep{Kobayashi+04, Guillochon+09}, although we note that \citet{Kobayashi+04}, using a one-dimensional mesh code, found results in qualitative agreement with BL08.  On the other hand, \citet{Guillochon+09} found a vastly lower degree ($\rho_{\rm c}\approx 4\rho_*$) of central compression for a $\beta=7$ event than what is predicted analytically, by the affine model, or by one-dimensional hydrodynamic simulations, with the difference attributed to a combination of insufficient midplane resolution and the physical, three-dimensional effects discussed in \S 5.  However, we have shown that these multidimensional effects are unlikely to be large due to the proximity of the sonic point to the region of maximum compression.

During the completion of this paper, an independent numerical study \citep{GuillochonRamirez12} was posted which found an approximate constancy in $\Delta\epsilon$ for $1\lesssim \beta \lesssim 4$, in the course of a thorough investigation of low-$\beta$ TDEs.  This result is broadly compatible with our paper, although further hydrodynamical simulations will be required to explore high $\beta$ values, and to test our prediction of desynchronized collapse for rapidly rotating, misaligned stars (which could restore a $\beta$ dependence to $\Delta \epsilon$, for deeply plunging disruptions).  Like \citet{GuillochonRamirez12}, we argue that the assumption of frozen-in debris energy is made much more valid by assuming the energy freezes in when internal forces become negligible at $R=R_{\rm t}$, not at later times.

We note that the free solutions are of course an approximation to the physical reality of tidal disruption, and neglect a number of important physical effects.  Some areas in particular need of further clarification by hydrodynamical simulation are the velocity and shape perturbations induced on the star by the nonlinear hydrodynamics of tidal disruption at $R\approx R_{\rm t}$.  Although BL08 found these effects to generally be negligble for Phase II and III of a TDE, their simulations were one-dimensional and it is conceivable that in three dimensions the picture may change.  The free solutions also fail to capture GR effects relevant for large $\beta$ values, although we have argued based on our own analysis and past literature that these effects do not qualitatively change our conclusions.  Furthermore, we have not considered the effects of SMBH spin, which if misaligned with the orbital plane may be able to desynchronize stellar collapse.

Despite these limitations, the free solutions are a simple yet powerful method for considering the tidal free fall of disrupted bodies.  Although we have focused on the disruption of main sequence stars by supermassive black holes, we stress that much of our analysis is general and applies equally well to other scenarios, such as tidal disruption of white dwarfs by intermediate mass black holes, or of planets or asteroids by more compact bodies.  In all of these scenarios, the decoupling of vertical from in-plane motion will strongly compress the disrupted object, often by many orders of magnitude.  In future work, we aim to apply the free solutions to these alternative physical regimes, as well as to internal motions such as pre-disruption rotation or post-disruption shock formation.

\section*{Acknowledgments}
NS would like to thank Michael Kesden and Bence Kocsis for helpful discussions.  NS and AL were supported in part by NSF Grant No. AST-0907890 and NASA Grants No. NNX08AL43G and No. NNA09DB30A.  RS was supported in part by a Radcliffe Fellowship, a Guggenheim Fellowship, and an ERC Grant.

\appendix

\section{Principal Axes of Free Solutions}

Using the in-plane free solutions, we find that the initial phases of the principal axes at $f=f_{\rm c}$ are given by
\begin{equation}
\cos \theta_{\rm ex}= \pm \sqrt{\frac{1}{2} \pm \frac{j}{2\sqrt{j^2+4k^2}}}
\end{equation}
where 
\begin{align}\label{jks}
j=&-\frac{16}{25}\beta+\frac{148}{25}+\frac{1312}{25\beta}-\frac{352}{5\beta^2}\\
 &+\sqrt{1-\beta^{-1}}\left(-\frac{16}{25}\beta+\frac{788}{25}-\frac{2208}{25\beta}+\frac{32}{5\beta^2}\right)\notag \\ 
k=&\sqrt{\beta}\left( \frac{72}{25}-\frac{936}{25\beta}+\frac{1144}{25\beta^2}-\frac{16}{5\beta^3} \right)\notag\\
&+\sqrt{\beta -1} \left( \frac{72}{25}+\frac{216}{25\beta}-\frac{176}{5\beta^2} \right) \notag
\end{align}
are exact expressions.  Note that this gives us 8 possible solutions for $\theta_{\rm ex}$.  While 4 are spurious, the other 4 of these are valid, with each minimum (maximum) in $r_{\rm H}$ possessing an equal minimum (maximum) $180^{\circ}$ around the star's center of mass.  In particular, for $\sin \theta_0 >0$, 
\begin{equation}
\cos \theta_{\rm min}= (-1)^{p} \sqrt{\frac{1}{2} +(-1)^{q} \frac{j}{2\sqrt{j^2+4k^2}}}
\end{equation}
and
\begin{equation}
\cos \theta_{\rm max}= (-1)^{P} \sqrt{\frac{1}{2} +(-1)^{Q} \frac{j}{2\sqrt{j^2+4k^2}}}.
\end{equation}

Here the behavior of $\theta_{\rm min}$ and $\theta_{\rm max}$ is piecewise with respect to $\beta$, due to zeros of $j$ and $k$.  We can describe this behavior for all $\beta>1$ by setting
\begin{equation}
\{p, q, P, Q\}=
\begin{cases}
\{1, 2, 2, 1 \}, \beta<1.073 \\
\{2, 1, 1, 2 \}, 1.073\le \beta < 4.944 \\
\{1, 2, 2, 1 \}, 4.944 \le \beta .
\end{cases}
\end{equation}

To gain better intuition for the geometry of these principal axes, we can Taylor expand $r_{\rm long}=r_{\rm H}(\theta_{\rm max})$ and $r_{\rm short}=r_{\rm H}(\theta_{\rm min})$ in the limit of large $\beta$:
\begin{align}
r_{\rm long}=& \frac{4}{5}\beta^{1/2}+\frac{22}{5}\beta^{-1/2}+O(\beta^{-3/2}) \label{rLongLim} \\
r_{\rm short}=& 2\beta^{-1/2}-\frac{23}{2}\beta^{-3/2}+O(\beta^{-5/2}). \label{rShortLim}
\end{align}
These results are poorly convergent for $\beta<10$, but describe the exact solutions well above this threshold.  We are now in a position to derive approximate formulae for the angles presented in Fig. \ref{angles}.  By expanding the numerator and denominator of $\tan\psi_{\rm c}=-y_{\rm H}(\theta_{\rm max})/x_{\rm H}(\theta_{\rm max})$ we find
\begin{equation}
\tan\psi_{\rm c} \approx \frac{16\beta^{1/2}+86\beta^{-1/2}}{8+185\beta^{-1}}.
\end{equation}
Using trigonometric identities and the Keplerian expression $\tan\Upsilon_{\rm c}=\sqrt{\beta}+\sqrt{\beta-1}\approx2\beta^{1/2}$ ($\Upsilon_{\rm c}$ is the angle between the $\hat{x}$ and a parabolic orbit's velocity vector at $f=f_{\rm c}$), we find the misalignment angle
\begin{equation}
\tan \nu_{\rm c}\approx\frac{284}{32\beta^{3/2}+180\beta^{1/2}+185\beta^{-1/2}}.
\end{equation}
We have defined $\nu_{\rm c}$ as the (positive) angle between the center of mass velocity and the long principal axis at $f=f_{\rm c}$.  It asymptotes to $0^{\circ}$ as $\beta$ goes to $\infty$.  


\begin{thebibliography}{99}

\bibitem[Bade et 
al.(1996)]{Bade+96} Bade, N., Komossa, S., \& Dahlem, M.\ 1996, A\&A, 309, L35 

\bibitem[Komossa 
\& Greiner(1999)]{KomossaGreiner99} Komossa, S., \& Greiner, J.\ 1999, A\&A, 349, L45 

\bibitem[Gezari et al.(2003)]{Gezari+03} Gezari, S., Halpern, 
J.~P., Komossa, S., Grupe, D., \& Leighly, K.~M.\ 2003, ApJ, 592, 42 

\bibitem[Gezari et al.(2006)]{Gezari+06} Gezari, S., Martin, 
D.~C., Milliard, B., et al.\ 2006, ApJL, 653, L25 

\bibitem[Gezari et al.(2008)]{Gezari+08} Gezari, S., Basa, S., 
Martin, D.~C., et al.\ 2008, ApJ, 676, 944 

\bibitem[Gezari et al.(2009)]{Gezari+09} Gezari, S., Heckman, T., 
Cenko, S.~B., et al.\ 2009, ApJ, 698, 1367 

\bibitem[van Velzen et al.(2011)]{vanVelzen+11} van Velzen, S., 
Farrar, G.~R., Gezari, S., et al.\ 2011, ApJ, 741, 73 

\bibitem[Cenko et al.(2012)]{Cenko+12} Cenko, S.~B., Bloom, 
J.~S., Kulkarni, S.~R., et al.\ 2012, MNRAS, 420, 2684 

\bibitem[Gezari et al.(2012)]{Gezari+12} Gezari, S., Chornock, 
R., Rest, A., et al.\ 2012, Nature, 485, 217 

\bibitem[Donley et al.(2002)]{Donley+02} Donley, J.~L., Brandt, 
W.~N., Eracleous, M., \& Boller, T.\ 2002, AJ, 124, 1308 

\bibitem[Levan et al.(2011)]{Levan+11} Levan, A.~J., Tanvir, 
N.~R., Cenko, S.~B., et al.\ 2011, Science, 333, 199 

\bibitem[Bloom et al.(2011)]{Bloom+11} Bloom, J.~S., Giannios, 
D., Metzger, B.~D., et al.\ 2011, Science, 333, 203 

\bibitem[Zauderer et al.(2011)]{Zauderer+11} Zauderer, B.~A., 
Berger, E., Soderberg, A.~M., et al.\ 2011, Nature, 476, 425 

\bibitem[Cenko et al.(2012)]{Cenko+11} Cenko, S.~B., Krimm, 
H.~A., Horesh, A., et al.\ 2012, ApJ, 753, 77 

\bibitem[Frank \& Rees(1976)]{FrankRees76} Frank, J., \& Rees, M.~J.\ 1976, MNRAS, 176, 633 

\bibitem[Lightman 
\& Shapiro(1977)]{LightmanShapiro77} Lightman, A.~P., \& Shapiro, S.~L.\ 1977, ApJ, 211, 244 

\bibitem[Cohn \& Kulsrud(1978)]{CohnKulsrud78} Cohn, H., \& Kulsrud, R.~M.\ 1978, ApJ, 226, 1087 

\bibitem[Magorrian 
\& Tremaine(1999)]{MagorrianTremaine99} Magorrian, J., \& Tremaine, S.\ 1999, MNRAS, 309, 447 

\bibitem[Wang 
\& Merritt(2004)]{WangMerritt04} Wang, J., \& Merritt, D.\ 2004, ApJ, 600, 149 

\bibitem[Merritt 
\& Poon(2004)]{MerrittPoon04} Merritt, D., \& Poon, M.~Y.\ 2004, ApJ, 606, 788 

\bibitem[Perets et al.(2007)]{Perets+07} Perets, H.~B., Hopman, 
C., \& Alexander, T.\ 2007, ApJ, 656, 709 

\bibitem[Ivanov et al.(2005)]{Ivanov+05} Ivanov, P.~B., Polnarev, 
A.~G., \& Saha, P.\ 2005, MNRAS, 358, 1361 

\bibitem[Chen et al.(2009)]{Chen+09} Chen, X., Madau, P., 
Sesana, A., \& Liu, F.~K.\ 2009, ApJL, 697, L149 

\bibitem[Chen et al.(2011)]{Chen+11} Chen, X., Sesana, A., 
Madau, P., \& Liu, F.~K.\ 2011, ApJ, 729, 13 

\bibitem[Wegg 
\& Nate Bode(2011)]{WeggBode11} Wegg, C., \& Nate Bode, J.\ 2011, ApJL, 738, L8 

\bibitem[Stone 
\& Loeb(2011)]{StoneLoeb11} Stone, N., \& Loeb, A.\ 2011, MNRAS, 412, 75 

\bibitem[Rees(1988)]{Rees88} Rees, M.~J.\ 1988, Nature, 333, 523 

\bibitem[Phinney(1989)]{Phinney89} Phinney, E.~S.\ 1989, The 
Center of the Galaxy, 136, 543 

\bibitem[Ulmer(1999)]{Ulmer99} Ulmer, A.\ 1999, ApJ, 514, 180 

\bibitem[Strubbe 
\& Quataert(2009)]{StrubbeQuataert09} Strubbe, L.~E., \& Quataert, E.\ 2009, MNRAS, 400, 2070 

\bibitem[Nolthenius 
\& Katz(1982)]{NoltheniusKatz82} Nolthenius, R.~A., \& Katz, J.~I.\ 1982, ApJ, 263, 377 

\bibitem[Evans 
\& Kochanek(1989)]{EvansKochanek89} Evans, C.~R., \& Kochanek, C.~S.\ 1989, ApJL, 346, L13

\bibitem[Laguna et al.(1993)]{Laguna+93} Laguna, P., Miller, 
W.~A., Zurek, W.~H., \& Davies, M.~B.\ 1993, ApJL, 410, L83 

\bibitem[Lodato et al.(2009)]{Lodato+09} Lodato, G., King, A.~R., 
\& Pringle, J.~E.\ 2009, MNRAS, 392, 332 

\bibitem[Guillochon et al.(2009)]{Guillochon+09} Guillochon, J., 
Ramirez-Ruiz, E., Rosswog, S., \& Kasen, D.\ 2009, ApJ, 705, 844 

\bibitem[Guillochon 
\& Ramirez-Ruiz(2012)]{GuillochonRamirez12} Guillochon, J., \& Ramirez-Ruiz, E.\ 2012, arXiv:1206.2350 

\bibitem[Kasen 
\& Ramirez-Ruiz(2010)]{KasenRamirez10} Kasen, D., \& Ramirez-Ruiz, E.\ 2010, ApJ, 714, 155 

\bibitem[Strubbe 
\& Quataert(2011)]{StrubbeQuataert11} Strubbe, L.~E., \& Quataert, E.\ 2011, MNRAS, 415, 168 

\bibitem[Kochanek(1994)]{Kochanek94} Kochanek, C.~S.\ 1994, ApJ, 
422, 508

\bibitem[Montesinos Armijo 
\& de Freitas Pacheco(2011)]{ArmijoPacheco11} Montesinos Armijo, M., \& de Freitas Pacheco, J.~A.\ 2011, ApJ, 736, 126 

\bibitem[Lodato 
\& Rossi(2011)]{LodatoRossi11} Lodato, G., \& Rossi, E.~M.\ 2011, MNRAS, 410, 359 

\bibitem[Loeb 
\& Ulmer(1997)]{LoebUlmer97} Loeb, A., \& Ulmer, A.\ 1997, ApJ, 489, 573 

\bibitem[Carter 
\& Luminet(1983)]{CarterLuminet83} Carter, B., \& Luminet, J.-P.\ 1983, A\&A, 121, 97 

\bibitem[Carter 
\& Luminet(1985)]{CarterLuminet85} Carter, B., \& Luminet, J.~P.\ 1985, MNRAS, 212, 23 

\bibitem[Luminet 
\& Carter(1986)]{LuminetCarter86} Luminet, J.-P., \& Carter, B.\ 1986, ApJS, 61, 219 

\bibitem[Luminet 
\& Marck(1985)]{LuminetMarck85} Luminet, J.-P., \& Marck, J.-A.\ 1985,MNRAS, 212, 57 

\bibitem[Luminet 
\& Pichon(1989)]{LuminetPichon89} Luminet, J.-P., \& Pichon, B.\ 1989, A\&A, 209, 85 

\bibitem[Kobayashi et al.(2004)]{Kobayashi+04} Kobayashi, S., 
Laguna, P., Phinney, E.~S., \& M{\'e}sz{\'a}ros, P.\ 2004, ApJ, 615, 855 

\bibitem[Diener et al.(1995)]{Diener+95} Diener, P., Kosovichev, 
A.~G., Kotok, E.~V., Novikov, I.~D., 
\& Pethick, C.~J.\ 1995, MNRAS, 275, 498 

\bibitem[Kesden(2011)]{Kesden11} Kesden, M.\ 2011, PRD, 83, 
104011 

\bibitem[Sari et al.(2010)]{Sari+10} Sari, R., Kobayashi, S., 
\& Rossi, E.~M.\ 2010, ApJ, 708, 605 

\bibitem[Brassart 
\& Luminet(2008)]{BrassartLuminet08} Brassart, M., \& Luminet, J.-P.\ 2008, A\&A, 481, 259 

\bibitem[Brassart 
\& Luminet(2010)]{BrassartLuminet10} Brassart, M., \& Luminet, J.-P.\ 2010, A\&A, 511, A80 

\bibitem[Blanchet(2006)]{Blanchet06} Blanchet, L.\ 2006, Living 
Reviews in Relativity, 9, 4 

\bibitem[Kesden(2012)]{Kesden12} Kesden, M.\ 2012, 
arXiv:1207.6401 

\bibitem[Rosswog et al.(2009)]{RosswogRR09} Rosswog, S., 
Ramirez-Ruiz, E., \& Hix, W.~R.\ 2009, ApJ, 695, 404 

\bibitem[Haas et al.(2012)]{Haas+12} Haas, R., Shcherbakov, 
R.~V., Bode, T., \& Laguna, P.\ 2012, ApJ, 749, 117 

\bibitem[Zalamea et al.(2010)]{Zalamea+12} Zalamea, I., Menou, K., 
\& Beloborodov, A.~M.\ 2010, MNRAS, 409, L25 

\end{thebibliography}
\end{document}